\documentclass[12pt]{article}
\usepackage{graphicx}

\usepackage{epsf}
\def\beq{\begin{equation}}
\def\eeq{\end{equation}}
\def\bea{\begin{eqnarray}}
\def\eea{\end{eqnarray}}
\def\ve{\vert}
\def\vel{\left|}
\def\ver{\right|}
\def\nnb{\nonumber}
\def\ga{\left(}
\def\dr{\right)}
\def\aga{\left\{}
\def\adr{\right\}}
\def\rar{\rightarrow}
\def\nnb{\nonumber}
\def\la{\langle}
\def\ra{\rangle}
\def\ba{\begin{array}}
\def\ea{\end{array}}
\def\bea{\begin{eqnarray}}
\def\eea{\end{eqnarray}}

\def\ds{\displaystyle}
\def\ve{\vert}
\def\vel{\left|}
\def\ver{\right|}
\def\nnb{\nonumber}
\def\ga{\left(}
\def\dr{\right)}
\def\aga{\left\{}
\def\adr{\right\}}
\def\rar{\rightarrow}
\def\nnb{\nonumber}
\def\la{\langle}
\def\ra{\rangle}
\def\lla{\left<}
\def\rra{\right>}
\def\simlt{\stackrel{<}{{}_\sim}}

\begin{document}
\def\beq{\begin{equation}}
\def\eeq{\end{equation}}
\def\bea{\begin{eqnarray}}
\def\eea{\end{eqnarray}}
\def\ve{\vert}
\def\vel{\left|}
\def\ver{\right|}
\def\nnb{\nonumber}
\def\ga{\left(}
\def\dr{\right)}
\def\aga{\left\{}
\def\adr{\right\}}
\def\rar{\rightarrow}
\def\nnb{\nonumber}
\def\la{\langle}
\def\ra{\rangle}
\def\lla{\left<}
\def\rra{\right>}
\def\ba{\begin{array}}
\def\ea{\end{array}}
\def\Bphi{$B_s \rar \phi \ell^+ \ell^-$}
\def\Bphimu{$B_s \rar \phi \mu^+ \mu^-$}
\def\Bphitau{$B_s \rar \phi \tau^+ \tau^-$}
\def\tepm{$B \rar K \mu^+ \mu^-$}
\def\tept{$B \rar K \tau^+ \tau^-$}
\def\ds{\displaystyle}
\title{ {\Large {\bf
Analysis of {$B_s \rar \phi \ell^+ \ell^-$} decay with new physics
effects} } }



\author
{{\small U. O. Yilmaz \thanks {e-mail: uoyilmaz@mersin.edu.tr}}\\
            \\
{\small  The J. Stefan Institute, Jamova 39, P. O. Box 300, 1001 Ljubljana, Slovenia} \\
{\small and} \\
{\small Physics Department, Mersin University 33343 Ciftlikkoy Mersin, Turkey} \\
}
\date { }

\begin{titlepage}
\maketitle \thispagestyle{empty}

\begin{abstract}
The rare \Bphi decay is investigated  by using the most general
model independent effective Hamiltonian for $\ell= \mu, \tau$.
The calculated $Br(B_s \rar \phi \mu^+ \mu^- ) = 1.92 \times 10^{-6}$
is in consistent with the experimental upper bound.
The dependencies of the branching ratios and polarization asymmetries of leptons
and combined lepton-antilepton asymmetries
on the new Wilson coefficients are presented. The analysis shows
that the branching ratios and the lepton polarization asymmetries
are very sensitive to the scalar and tensor type interactions. The
results obtained in this work will be very useful in searching new physics beyond the
standard model.
\end{abstract}

\end{titlepage}

\section{Introduction}
In searching the new physics, the analysis of rare B decays induced
by flavor changing neutral current (FCNC) can play an important role.
Since the observation of exclusive $B\rar K^* \gamma$ \cite{CLEO} decay,
which sitimulated the works in this area,
there have been increasing number of investigations of new physics
both in theoretical and experimental side, induced by FCNC $b\rar s, d$ \cite{Artuso}-\cite{Ali05}  transitions.
These processes occur at loop level in the SM and are very sensitive to the gauge structure
and various extensions of the SM so they can give useful
information on parameters of the SM, testing its predictions at loop level
and also probing new physics.
The new physics effects in rare B meson decays can indicate itself through
new contributions to the Wilson coefficients that are already present in the SM, and through
the new operators in the effective Hamiltonian which are absent in the SM. In this
work we use a most general model-independent effective Hamiltonian that combines both
approaches and contains the scalar and tensor type interactions as well as the vector
types.

In this work we consider the \Bphi decay induced by FCNC  $b\rar s
\ell^+ \ell^-$ transition at quark level. This semileptonic decay
is one of the  suitable tools of investigating new physics, via
calculating many observables since it occurs only at loop level in
the SM. However, being an exclusive decay, its theoretical
investigation may be more difficult than that of corresponding
inclusive decays, although experimentally the situation is in
contrary. This difficulty is because of requiring additional
knowledge on form factors, the matrix elements of the Hamiltonian
between initial and final meson states. This is related to the
nonperturbative sector of QCD and can be solved in the framework
of nonperturbative approaches. For \Bphi decay, the matrix
elements of the effective Hamiltonian between the initial and
final states have been calculated in the framework of different
approaches, such as light cone sum rules \cite {Ball1998,
Ball2005, Wu2006} and in different quark models; relativistic
constitute quark model \cite{Melikov2000}, constituent quark model
\cite{Deandrea} and light front quark model \cite{Geng2003}.

In establishing new physics beyond the SM, measurement of the lepton
polarization is an useful and efficient way. Besides different decay modes, polarization properties of
exclusive semileptonic decay modes have been studied
both in the SM \cite {Geng96}-\cite{Aliev2000} and beyond \cite{Aliev01}-\cite{uoyilmaz07}.
In this work we study the branching ratio and lepton
polarization asymmetries in the exclusive \Bphi decay using
the model independent general effective Hamiltonian. This decay has already been studied in the SM \cite{Geng2003},
two Higgs doublet model \cite {Gursevil2002} and in a universal
extra dimension scenario \cite {Mohanta2007}.

On the experimental side, the first limit on branching fraction was stated by CDF Collaboration \cite{CDF2002}.
DO Collaboration \cite{DO2006} has reported an upper
limit on the branching ratio for $Br(B_s \rar \phi \mu^+ \mu^- )
<4.1 \times 10^{-6}$ and recently, another upper bound $Br(B_s \rar
\phi \mu^+ \mu^-) <6.0(5.0) \times 10^{-6}$ at $95(90) \%$ C.L.  by CDF Collaboration
\cite{CDF2008}. At LHCb for a nominal year of data taking ($2 fb^{-1}$)
about $1340\pm250$ annual yield of \Bphimu\,  is expected over
$10^{12}$ $b\bar b$ events \cite{Tayduganov2007}-\cite{Altarelli}.
At Atlas Experiment, $ \sim 600$ \Bphimu
can be reconstructed over a three-year period with $30 fb^{-1}$.
These experiments and also the running B factories encourage
the study of rare B meson decays, in our case \Bphi decay.

This work is organized as follows. In section 2, we drive the model
independent expressions for the longitudinal, transversal and normal
polarizations of leptons and combined lepton-antilepton
polarizations starting from the quark level process and using
appropriate form factors for \Bphi decay. In section 3, we present
our numerical results and the dependence of branching ratios and polarizations on new Wilson
coefficients for $\mu, \tau$.

\section{Effective Hamiltonian and lepton polarizations}
The quark level transition of the \Bphi decay is described by  $b
\rar s \ell^+ \ell^- $ in the standard effective Hamiltonian
approach, and can be written in term of twelve four--Fermi interactions,
including all possible terms calculated independent of any
models,
as follows \cite{Fukae2000},
\bea
\label{effH} {\cal H}_{eff} &=&
\frac{G\alpha}{\sqrt{2} \pi}
 V_{ts}V_{tb}^\ast
    \Bigg\{ C_{SL} \, \bar s i \sigma_{\mu\nu} \frac{q^\nu}{q^2}\, L \,b
    \, \bar \ell \gamma^\mu \ell + C_{BR}\, \bar s i \sigma_{\mu\nu}
    \frac{q^\nu}{q^2} \,R\, b \, \bar \ell \gamma^\mu \ell \nnb \\
&+&C_{LL}^{tot}\, \bar s_L \gamma_\mu b_L \,\bar \ell_L \gamma^\mu
    \ell_L + C_{LR}^{tot} \,\bar s_L \gamma_\mu b_L \, \bar \ell_R
    \gamma^\mu \ell_R +
    C_{RL} \,\bar s_R \gamma_\mu b_R \,\bar \ell_L \gamma^\mu \ell_L \nnb \\
&+&C_{RR} \,\bar s_R \gamma_\mu b_R \, \bar \ell_R \gamma^\mu \ell_R
    + C_{LRLR} \, \bar s_L b_R \,\bar \ell_L \ell_R +
    C_{RLLR} \,\bar s_R b_L \,\bar \ell_L \ell_R \\
&+&C_{LRRL} \,\bar s_L b_R \,\bar \ell_R \ell_L + C_{RLRL} \,\bar
    s_R b_L \,\bar \ell_R \ell_L+
    C_T\, \bar s \sigma_{\mu\nu} b \,\bar \ell \sigma^{\mu\nu}\ell \nnb \\
&+&i C_{TE}\,\epsilon^{\mu\nu\alpha\beta} \bar s \sigma_{\mu\nu} b
    \, \bar \ell \sigma_{\alpha\beta} \ell  \Bigg\}~, \nnb
\eea
where $L = 1-\gamma_5/2$ and $R = 1+\gamma_5/2$ are the chiral
projection operators and $C_X$ are the coefficients of the
four--Fermi interactions. The coefficients $C_{SL}$ and $C_{BR}$ are
the nonlocal Fermi interactions and their correspondence in the SM
are $-2 m_s C_7^{eff}$ and $-2 m_b C_7^{eff}$, respectively. The
terms with coefficients $C_{LL}$, $C_{LR}$, $C_{RL}$ and $C_{RR}$
are the vector interactions. Two of them are written as $C_{LL}^{tot} = C_9^{eff} - C_{10} +
C_{LL}$ and $C_{LR}^{tot} = C_9^{eff} + C_{10} + C_{LR}$. So these terms describe the sum of the contributions
from the SM and the new physics. The terms with coefficients
$C_{LRLR}$, $C_{RLLR}$, $C_{LRRL}$ and $C_{RLRL}$ describe the
scalar type and $C_T$ and $C_{TE}$ describe the tensor type
interactions.

Having given the general form of four--Fermi interaction for the $b
\rar s \ell^+ \ell^- $ transition, we now need to calculate the
matrix element for the \Bphi decay in order to calculate the decay
amplitude. These transition matrix elements can be written in terms
of invariant form factors over $B_s$ and $\phi$ in the following
form \cite {Ball1998} together with \cite{Aliev2005},
\bea
 \lefteqn{ \label{ilk} \lla
    \phi(p_{\phi},\varepsilon) \vel \bar s \gamma_\mu
    (1 \pm \gamma_5) b \ver B_s(p_{B_s}) \rra =} \nnb \\
&&- \epsilon_{\mu\nu\alpha\beta} \varepsilon^{\ast\nu}
    p_{\phi}^\alpha q^\beta \frac{2 V(q^2)}{m_{B_s}+m_{\phi}}
    \pm
    i \varepsilon_\mu^\ast (m_{B_s}+m_{\phi})
    A_1(q^2) \\
&&\mp i (p_{B_s} + p_{\phi})_\mu (\varepsilon^\ast q)
    \frac{A_2(q^2)}{m_{B_s}+m_{\phi}}
    \mp i q_\mu (\varepsilon^\ast q)\frac{2 m_{\phi}}{q^2} [A_3(q^2) - A_0(q^2)] ~,  \nnb \\
    \nnb \\
\lefteqn{ \label{iki} \lla \phi(p_{\phi},\varepsilon) \vel
    \bar s i \sigma_{\mu\nu} q^\nu
    (1 \pm \gamma_5) b \ver B_s(p_{B_s}) \rra =} \nnb \\
&&2 \epsilon_{\mu\nu\alpha\beta} \varepsilon^{\ast\nu}
    p_{\phi}^\alpha q^\beta T_1(q^2) \pm  i \left[
    \varepsilon_\mu^\ast (m_{B_s}^2-m_{\phi}^2) -
    (p_{B_s} + p_{\phi})_\mu (\varepsilon^\ast q) \right] T_2(q^2) \\
&&\pm  i (\varepsilon^\ast q) \left[ q_\mu - (p_{B_s} +
    p_{\phi})_\mu \frac{q^2}{m_{B_s}^2-m_{\phi}^2} \right] T_3 (q^2)~, \nnb \\  \nnb \\
\lefteqn{ \label{ucc} \lla \phi (p_{\phi},\varepsilon) \vel
    \bar s \sigma_{\mu\nu}
     b \ver B_s(p_{B_s}) \rra =} \nnb \\
&&i \epsilon_{\mu\nu\alpha\beta}  \Bigg[ -  T_1(q^2)
    {\varepsilon^\ast}^\alpha (p_{B_s} + p_{\phi})^\beta +
    \frac{1}{q^2} (m_{B_s}^2-m_{\phi}^2) {\varepsilon^\ast}^\alpha
    q^\beta [T_1(q^2)-T_2(q^2)]\\
&&- \frac{2}{q^2} \Bigg(T_1(q^2) - T_2(q^2) - \frac {q^2}
        {m_{B_s}^2-m_{\phi}^2} T_3 (q^2) \Bigg) (\varepsilon^\ast
    q) p_{\phi}^\alpha q^\beta \Bigg]~, \nnb \eea
and \bea \label{uc}\lla \phi(p_{\phi},\varepsilon) \vel \bar s (1
    \pm
    \gamma_5) b \ver B_s(p_{B_s}) \rra & =
    & \frac{1}{m_b} \Big[ \mp  2i m_{\phi} (\varepsilon^\ast q) A_0(q^2)
    \Big]~,
\eea
where $q = p_{B_s}-p_{\phi}$ is the momentum transfer and
$\varepsilon$ is the polarization vector of $\phi$ meson. The matrix
element in (\ref {uc}) is calculated by contracting both sides
of (\ref{ilk}) with $q^\mu$, using equation of motion and the
following relation \cite{Aliev2005}
\bea \label{AA3}  A_3(q^2) = \frac {m_{B_s} + m_{\phi}}
                        {2 m_{\phi}}  A_1(q^2)
                - \frac {m_{B_s} - m_{\phi}} {2 m_{\phi}} A_2(q^2). \nnb
\eea
In order to avoid kinematical singularity in the matrix element at
$q^2=0$, it is assumed that $A_0(0) = A_3(0)$ and $T_1(0)=T_2(0)$ \cite{Ball1998}.\\

After the definitions of the form factors, the matrix element of the
\Bphi decay can be written by using (\ref{effH})--(\ref{uc})
as,
\bea \lefteqn{ \label{had}
    {\cal M}($\Bphi$) =
    \frac {G \alpha}{4 \sqrt{2} \pi} V_{tb} V_{ts}^\ast }\nnb \\
&&\times \Bigg\{
    \bar \ell \gamma^\mu(1-\gamma_5) \ell \, \Big[
    -2 A_1 \epsilon_{\mu\nu\alpha\beta} \varepsilon^{\ast\nu}
    p_{\phi}^\alpha q^\beta
    -i B_1 \varepsilon_\mu^\ast
    + i B_2 (\varepsilon^\ast q) (p_{B_s}+p_{\phi})_\mu
    + i B_3 (\varepsilon^\ast q) q_\mu  \Big] \nnb \\
&&+ \bar \ell \gamma^\mu(1+\gamma_5) \ell \, \Big[
    -2 C_1 \epsilon_{\mu\nu\alpha\beta} \varepsilon^{\ast\nu}
    p_{\phi}^\alpha q^\beta
    -i D_1 \varepsilon_\mu^\ast
    + i D_2 (\varepsilon^\ast q) (p_{B_s}+p_{\phi})_\mu
    + i D_3 (\varepsilon^\ast q) q_\mu  \Big] \nnb \\
&&+\bar \ell (1-\gamma_5) \ell \Big[ i B_4 (\varepsilon^\ast
    q)\Big]
    + \bar \ell (1+\gamma_5) \ell \Big[ i B_5 (\varepsilon^\ast
    q)\Big]  \nnb \\
&&+4 \bar \ell \sigma^{\mu\nu}  \ell \Big( i C_T
\epsilon_{\mu\nu\alpha\beta}
    \Big) \Big[ -2 T_1 {\varepsilon^\ast}^\alpha (p_{B_s}+p_{\phi})^\beta +
    B_6 {\varepsilon^\ast}^\alpha q^\beta -
    B_7 (\varepsilon^\ast q) {p_{\phi}}^\alpha q^\beta \Big] \nnb \\
&&+16 C_{TE} \bar \ell \sigma_{\mu\nu}  \ell \Big[ -2 T_1
    {\varepsilon^\ast}^\mu (p_{B_s}+p_{\phi})^\nu  +B_6 {\varepsilon^\ast}^\mu q^\nu -
    B_7 (\varepsilon^\ast q) {p_{\phi}}^\mu q^\nu
    \Bigg\}~,
\eea
where
\bea \label{as}
A_1 &=& (C_{LL}^{tot} + C_{RL})
\frac{V}{m_{B_s}+m_{\phi}} -
    (C_{BR}+C_{SL}) \frac{T_1}{q^2} ~, \nnb \\
B_1 &=& (C_{LL}^{tot} - C_{RL}) (m_{B_s}-m_{\phi}) A_1-
    (C_{BR}-C_{SL}) (m_{B_s}^2-m_{\phi}^2)
    \frac{T_2}{q^2} ~, \nnb \\
B_2 &=& \frac{C_{LL}^{tot} - C_{RL}}{m_{B_s}+m_{\phi}} A_2 -
    (C_{BR}-C_{SL}) \frac{1}{q^2} \Big(T_2 + \frac {q^2}{m_{B_s}^2 - m_{\phi}^2} T_3 \Big) ~,
    \nnb \\
B_3 &=&  2 (C_{LL}^{tot} - C_{RL}) \frac{m_{\phi}} {q^2} (A_3 -A_0)+
     (C_{BR}-C_{SL}) \frac{ T_3}{q^2} ~, \nnb
    \\
C_1 &=& (C_{LR}^{tot} + C_{RR})
\frac{V}{m_{B_s}+m_{\phi}} -
    (C_{BR}+C_{SL}) \frac{T_1}{q^2} ~, \nnb \\
D_1 &=& (C_{LR}^{tot} - C_{RR}) (m_{B_s}-m_{\phi}) A_1-
    (C_{BR}-C_{SL}) (m_{B_s}^2-m_{\phi}^2)
    \frac{T_2}{q^2} ~, \nnb \\
D_2 &=& \frac{C_{LR}^{tot} - C_{RR}}{m_{B_s}+m_{\phi}} A_2 -
    (C_{BR}-C_{SL}) \frac{1}{q^2} \Big(T_2 + \frac {q^2}{m_{B_s}^2 - m_{\phi}^2} T_3 \Big) ~,
    \nnb \\
D_3 &=&  2 (C_{LR}^{tot} - C_{RR}) \frac{m_{\phi}} {q^2} (A_3 -A_0)+
     (C_{BR}-C_{SL}) \frac{ T_3}{q^2} ~, \nnb \\
B_4 &=& -  2( C_{LRRL} - C_{RLRL}) \frac{m_{\phi}}{m_b} A_0 ~,\nnb \\
B_5 &=& -  2 ( C_{LRLR} - C_{RLLR}) \frac{m_{\phi}}{m_b} A_0 \,,\nnb \\
B_6 &=&  (m_{B_s}^2-m_{\phi}^2) \frac{T_1 - T_2}{q^2} \, ,\nnb \\
B_7 &=& \frac{2}{q^2} \Big( T_1 - T_2 - \frac{q^2} {m_{B_s}^2 -
    m_{\phi}^2} T_3 \Big)  \,.  \eea
At this point, we would like to calculate the final lepton
polarizations for the \Bphi decay. In order to do this, we define
the orthogonal unit vector $S_{i}^{- \mu}$ in the rest frame of
$\ell^-$ and $S_{i}^{+ \mu}$ in the rest frame of $\ell^+$, for the
polarization of the leptons along the longitudinal ($L$),
transversal ($T$) and normal ($N$) directions.  Using the convention
of \cite{Fukae2000, Kruger96}, we can write
\bea \label{pol} S_L^{-\mu} &\equiv& (0,{\bf{e}}_L^{\,-}) =
        \ga 0,\frac{\bf{p}_-}{\vel \bf{p}_- \ver} \dr~, \nnb \\
S_N^{-\mu} &\equiv& (0,{\bf{e}}_N^{\,-}) = \ga 0,\frac{\bf{p} \times
    \bf{p}_-}
    {\vel \bf{p} \times \bf{p}_- \ver} \dr~, \nnb \\
S_T^{-\mu} &\equiv& (0,{\bf{e}}_T^{\,-}) =
    \ga 0, {\bf{e}}_N^{\,-} \times {\bf{e}}_L^{\,-} \dr~, \nnb \\
S_L^{+\mu} &\equiv& (0,{\bf{e}}_L^{\,+}) =
    \ga 0,\frac{\bf{p}_+}{\vel \bf{p}_+ \ver} \dr~, \nnb \\
S_N^{+\mu} &\equiv& (0,{\bf{e}}_N^{\,+}) = \ga 0,\frac{\bf{p} \times
    \bf{p}_+}
    {\vel \bf{p} \times \bf{p}_+ \ver} \dr~, \nnb \\
S_T^{+\mu} &\equiv& (0,{\bf{e}}_T^{\,+}) = \ga 0, {\bf{e}}_N^{\,+}
    \times {\bf{e}}_L^{\,+} \dr~,
 \eea
where $\bf{p}_\pm$ and $\bf{p}$ are the three momenta of
$\ell^{\pm}$ and $\phi$ meson in the center of mass (CM) frame of
the lepton pair system, respectively. The longitudinal unit vectors
$S_L^ {\pm}$ are boosted to CM frame of the lepton pair by Lorentz
transformation,
\bea \label{bs} S^{-\mu}_{L,\, CM} &=& \ga \frac{\vel \bf{p}_-
\ver}{m_\ell},
\frac{E_\ell \,\bf{p}_-}{m_\ell \vel \bf{p}_- \ver} \dr~, \nnb \\
S^{+\mu}_{L,\, CM} &=& \ga \frac{\vel \bf{p}_- \ver}{m_\ell}, -
\frac{E_\ell \, \bf{p}_-}{m_\ell \vel \bf{p}_- \ver} \dr~, \eea
while vectors of perpendicular directions are not changed by the
boost.

The differential decay rate of the \Bphi can be written in any spin
direction, as $\bf{n}^{\pm}$ being any spin direction of the
$\ell^{\pm}$, in the rest frame of lepton pair, in the following
form:
\bea
\label{ddr} \frac{d\Gamma(\bf{n}^{\pm})}{ds} = \frac{1}{2} \ga
\frac{d\Gamma}{ds}\dr_0 \Bigg[ 1 + \Bigg( P_L^{\pm}
{\bf{e}}_L^{\,\pm} + P_N^{\pm} {\bf{e}}_N^{\,\pm} + P_T^{\pm}
{\bf{e}}_T^{\,\pm} \Bigg) \cdot \bf{n}^{\pm} \Bigg]~, \eea
where $s=q^2/m_{B_s}^2$, the superscripts $^+$ and $^-$ respectively
correspond to $\ell^+$ and $\ell^-$ cases and $(d\Gamma / ds)_{0}$
corresponds to the unpolarized decay rate,
\bea \label{unp} \ga \frac{d \Gamma}{ds}\dr_0 &=& \frac{G^2
\alpha^2 m_{B_s}}{2^{14} \pi^5 }
    \vel V_{tb} V_{ts}^\ast \ver^2 \sqrt{\lambda} v \Delta
\eea
where $\Delta$ is given in Appendix A and $\lambda= 1 + r^2 + s^2 -2r-2s-2rs$
and lepton velocity is $v=\sqrt{1-{4m_\ell^2}/{s m_{B_s}^2}}$.

The polarizations $P^{\pm}_L$, $P^{\pm}_T$ and $P^{\pm}_N$ in (\ref{ddr}) are defined by the equation
\bea P_i^{\pm}(s) = \frac{\ds{\frac{d \Gamma}{ds}
                   ({\bf{n}}^{\pm}={\bf{e}}_i^{\,\pm}) -
                   \frac {d \Gamma}{ds}
                   ({\bf{n}}^{\pm}=-{\bf{e}}_i^{\,\pm})}}
             {\ds{\frac{d \Gamma}{ds}
             ({\bf{n}}^{\pm}={\bf{e}}_i^{\,\pm}) +
             \frac{d \Gamma}{ds}
             ({\bf{n}}^{\pm}=-{\bf{e}}_i^{\,\pm})}}~, \nnb
\eea
for $i=L,~N,~T$. Here, longitudinal and transversal asymmetries of the charged leptons $\ell^{\pm}$ in the
decay plane are $P^{\pm}_L$ and $P^{\pm}_T$, respectively, and the normal component
to both of them is $P^{\pm}_N$.

The main contribution to $P^{-}_L$ and $P^{-}_N$
is due to tensor interactions. In case of $P^{-}_T$, it also receives
considerable scaler contribution in additon to tensor effects.
The expressions for the lepton polarizations are given in the Appendix B.

From (\ref{plm})-(\ref{pnp}), it can be observed that for longitudinal
and normal polarizations, the difference between $\ell^+$ and
$\ell^-$ lepton asymmetries results from the scalar and tensor type
interactions. Similar situation takes place for transverse
polarization asymmetries in the $m_\ell \rar 0$ limit. From this, we
can conclude that their experimental study may provide   essential
information about new physics.

In searching new physics, the combined analysis of the lepton and
antilepton polarizations can be another useful source, since in the
SM $P_L^-+P_L^+=0$, $P_N^-+P_N^+= 0$ and $P_T^- - P_T^+ \approx 0$
\cite{Fukae2000}. Using (\ref{plm}) we obtain combined
longitudinal polarization,
the combined transversal polarization which is the difference of the lepton
and antilepton polarizations, from (\ref{ptm}) and
(\ref{ptp}) and finally the combined normal polarization,  from (\ref{pnm}) and
(\ref{pnp}). The explicit form of these asymmetries are also given in the Appendix B.

One should note from (\ref{lpl}) that in $P_L^-+P_L^+$ the SM contribution coming with $C_{BR}$,
$C_{SL}$, $C^{tot}_{LL}$ and $C^{tot}_{LR}$ terms, completely cancels. Any nonzero measurement of the value of $P_L^-+P_L^+$ in
future experiments, may be an evidence of the discovery of new
physics beyond the SM.
\section {Numerical analysis and discussion}
We will present our numerical analysis of the branching ratios and
polarizations and their dependencies on Wilson coefficients in a
series of figures, but before doing this, let us remark on a few
points.

The expressions of the lepton polarizations depend on both $s$ and
the new Wilson coefficients. It may not be experimentally easy to
study the polarizations depending on both quantities. So, by taking
the averaged forms over the allowed kinematical region, we eliminate
the dependency of the lepton polarizations on $s$. The averaged
lepton polarizations are defined by
\bea \label{av}
    \lla P_i \rra = \frac{\ds \int_{(2 m_\ell/m_{B_s})^2}^{(1-m_{\phi}/m_{B_s})^2}
                        P_i \frac{d{\cal B}}{ds} ds}
            {\ds \int_{(2 m_\ell/m_{B_s})^2}^{(1-m_{\phi}/m_{B_s})^2}
                     \frac{d{\cal B}}{ds} ds}~.
\eea

The input parameters we used in our numerical analysis are:
\begin{eqnarray}
\label{parameters}
 & & m_{B_s} =5.367 \, GeV \, , \,m_{\phi}=1.019 \,GeV\,\,\, m_b =4.8 \, GeV \, ,
        \,m_{\mu} =0.105 \, GeV \, , \,
        m_{\tau} =1.77 \, GeV \, , \nnb \\
& &  |V_{tb} V^*_{ts}|=0.0385 \, \, , \, \, \alpha^{-1}=129  \, \,
        ,G_F=1.17 \times 10^{-5}\, GeV^{-2}\, , \tau_{B_{s}}=1.425 \times 10^{-12} \, s \,
        . \nnb
\end{eqnarray}
The values of the individual Wilson coefficients that appear in the
SM at $\mu \sim m_b$ are listed in Table (\ref{table1}) and the
parameters that are not given here are taken from \cite{PDG}.

\begin{table}
        \begin{center}
        \begin{tabular}{|c|c|c|c|c|c|c|c|c|}
        \hline
        \multicolumn{1}{|c|}{ $C_1$}       &
        \multicolumn{1}{|c|}{ $C_2$}       &
        \multicolumn{1}{|c|}{ $C_3$}       &
        \multicolumn{1}{|c|}{ $C_4$}       &
        \multicolumn{1}{|c|}{ $C_5$}       &
        \multicolumn{1}{|c|}{ $C_6$}       &
        \multicolumn{1}{|c|}{ $C_7^{\rm eff}$}       &
        \multicolumn{1}{|c|}{ $C_9$}       &
                \multicolumn{1}{|c|}{$C_{10}$}      \\
        \hline
        $-0.248$ & $+1.107$ & $+0.011$ & $-0.026$ & $+0.007$ & $-0.031$ &
   $-0.313$ &   $+4.344$ &    $-4.624$       \\
        \hline
        \end{tabular}
        \end{center}
\caption{ Values of the SM Wilson coefficients at $\mu \sim m_b $
scale.\label{table1}}
\end{table}

The given $C^{eff}_9$ value in Table (\ref{table1}) corresponds only
to the short-distance contributions, but it should be noted that
$C^{eff}_9$ also receives long-distance contributions due to
conversion of the real $\bar{c}c$ into lepton pair $\ell^+ \ell^-$
and are usually absorbed into a redefinition of the short-distance
Wilson coefficients:
\begin{eqnarray}
C_9^{eff}(\mu)=C_9(\mu)+ Y(\mu)\,\, , \label{C9efftot}
\end{eqnarray}
where
\begin{eqnarray}
\label{EqY} Y(\mu)&=& Y_{reson}+ h(y,s) [ 3 C_1(\mu) + C_2(\mu) + 3
    C_3(\mu) + C_4(\mu) + 3 C_5(\mu) + C_6(\mu)] \nnb \\
    &-&
        \frac{1}{2} h(1, s) \left( 4 C_3(\mu) + 4 C_4(\mu)
        + 3 C_5(\mu) + C_6(\mu) \right)\nnb \\
&- &  \frac{1}{2} h(0,  s) \left[ C_3(\mu) + 3 C_4(\mu) \right] \nnb \\
&+& \frac{2}{9} \left( 3 C_3(\mu) + C_4(\mu) + 3 C_5(\mu) +
    C_6(\mu) \right)\,\, ,
\end{eqnarray}
with $y=m_c/m_b$, and the functions $h(y,s)$ arises from the one
loop contributions of the four quark operators $O_1$,...,$O_6$. The
explicit forms of them can be found in
\cite{Buras1995}-\cite{Misiakerratum}. Parametrization of the
resonance $\bar{c}c$ contribution, $Y_{reson}(s)$, given in
(\ref{EqY}) can be done by using a Breit-Wigner shape with
normalizations fixed by data given by \cite{AAli1991}
\begin{eqnarray}
Y_{reson}(s)&=&-\frac{3}{\alpha^2_{em}}\kappa \sum_{V_i=\psi_i}
\frac{\pi \Gamma(V_i\rightarrow \ell^+ \ell^-)m_{V_i}}{s
m^2_{B_{c}}-m_{V_i}+i m_{V_i}
\Gamma_{V_i}} \nonumber \\
&\times & [ 3 C_1(\mu) + C_2(\mu) + 3 C_3(\mu) + C_4(\mu) + 3
C_5(\mu) + C_6(\mu)]\, ,
 \label{Yresx}
\end{eqnarray}
where the phenomenological parameter $\kappa$ is taken as
$ 2.3$.

The new Wilson coefficients are the free parameters in this work,
but  it is possible to establish ranges out of experimentally
measured branching ratios of the semileptonic rare B-meson decays
\bea
BR (B \rar K \, \ell^+ \ell^-) & = & (4.8^{+1.0}_{-0.9}\pm 0.3) \times 10^{-7} \, \,\cite{Belle} ,\nnb \\
BR (B \rar K^* \, \mu^+ \mu^-) & = & (1.27^{+0.76}_{-0.61}\pm
                                                    0)\times 10^{-6}\, \,\cite{Babar} ,\nnb \eea
and also the upper bound of pure leptonic rare B-decays in the $B^0
\rar  \mu^+ \mu^-$ mode \cite{CDF05}:
\bea BR ( B^0 \rar  \mu^+ \mu^-) & \leq & 1.5 \times 10^{-7}  \, \,
.\nnb \eea
Compliant to this upper limit and the branching ratios for the
semileptonic rare B-decays, in this work we take all new Wilson
coefficients as real and varying in the region $-4\leq C_X\leq 4$.

The numerical values of the form factors that we used in this work
are the results of \cite{Ball1998}. The form factors are calculated
in light cone sum rule approach and include the radiative
and higher twist corrections and SU(3) breaking effects.
The $q^2$ dependencies of the form factors in three parameter fit
are given as
\begin{eqnarray}
F(q^2) = \frac{F(0)}{ 1-a s + b s^2}~, \nnb
\end{eqnarray}
where the values of parameters $F(0)$, $a$ and $b$ for the $B_s \rar
\phi$ decay are listed in Table 2.
\\
\begin{table}[h]
\renewcommand{\arraystretch}{1.5}
\addtolength{\arraycolsep}{3pt}
$$
\begin{array}{|l|c|c|c|c|c|c|c|}
\hline & A_0 & A_1 & A_2& V & T_1 & T_2 & T_3 \\ \hline
F(0) &
    \phantom{-} 0.382  & 0.296 & 0.255 & 0.433 & 0.348 & 0.348 & 0.254\\
a    &
   \phantom{-} 1.77  & 0.87 & 1.55 & 1.75 & 1.820 & 0.70 & 1.52 \\
b    &
     \phantom{-} 0.856 & -0.061 & 0.513 & 0.736 & 0.825 & -0.315 & 0.377 \\ \hline
\end{array}
$$
\caption{Light cone sum rule approach $B_s\rar \phi$ meson decay
form factors in a three parameters fit, including radiative and higher twist corrections and SU(3) breaking effects.}
\renewcommand{\arraystretch}{1}
\addtolength{\arraycolsep}{-3pt}
\end{table}
\\
Before discussing the figures including the results of our analysis,
we would like to give our SM predictions for the longitudinal,
transverse and the normal components of the lepton polarizations for
\Bphi decay for the $\mu$ ($\tau$) channel for reference:
\bea \label{smpre}
<P^-_{L}>  & =  & 0.8373 \, (0.4299) \, ,\nnb \\
<P^-_{T}>  & =  & 0.0025 \, (0.0498 ) \, ,\nnb \\
<P^-_{N}>  & =  & 0.0013 \, (0.0214) \, .\nnb \eea
The $<P^-_{L}>$ value is in consistent with $<P^-_{L}>_{\mu(\tau)}=
-0.81 (-0.49)$, respectively \cite{Geng2003}.
(The opposite sign is caused by the definition of the form factors.)

\begin{figure}
\centering
\includegraphics[width=5in]{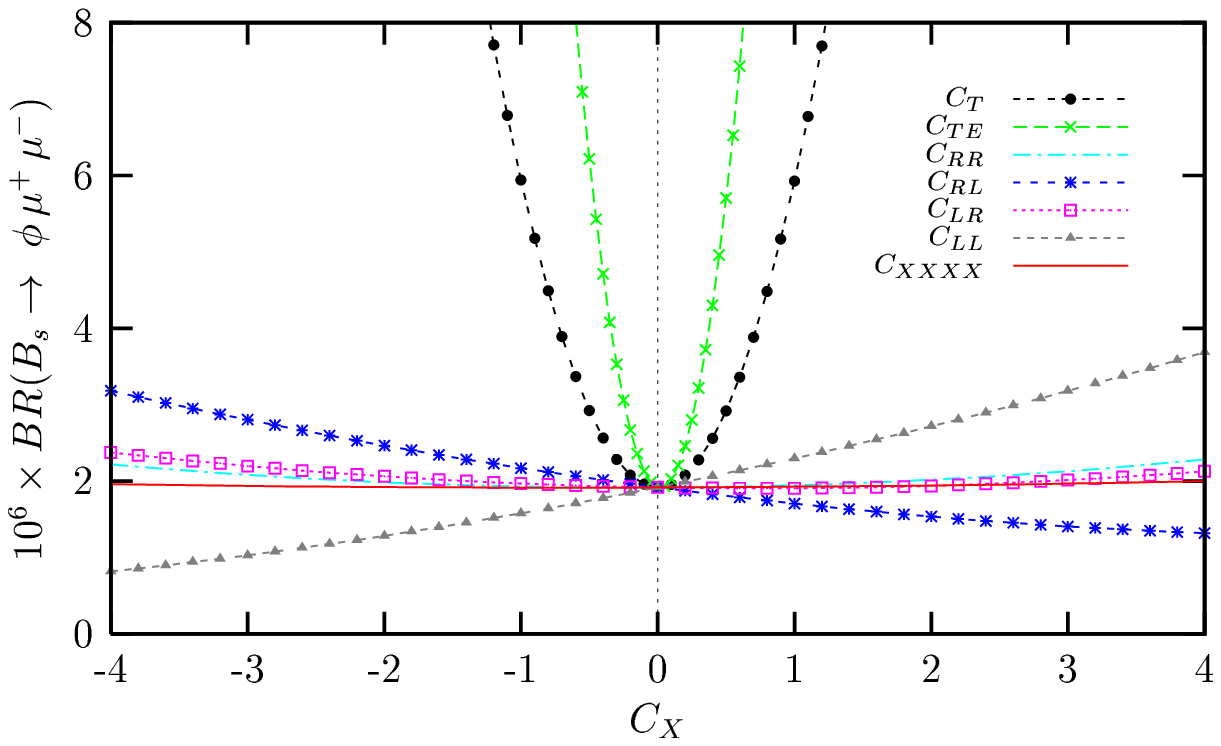}
\caption{The dependence of the integrated branching ratio for the
$B_s \rar \phi \, \mu^+ \mu^-$ decay on the new Wilson coefficients.
\label{f1}}
\end{figure}

\begin{figure}
\centering
\includegraphics[width=5in]{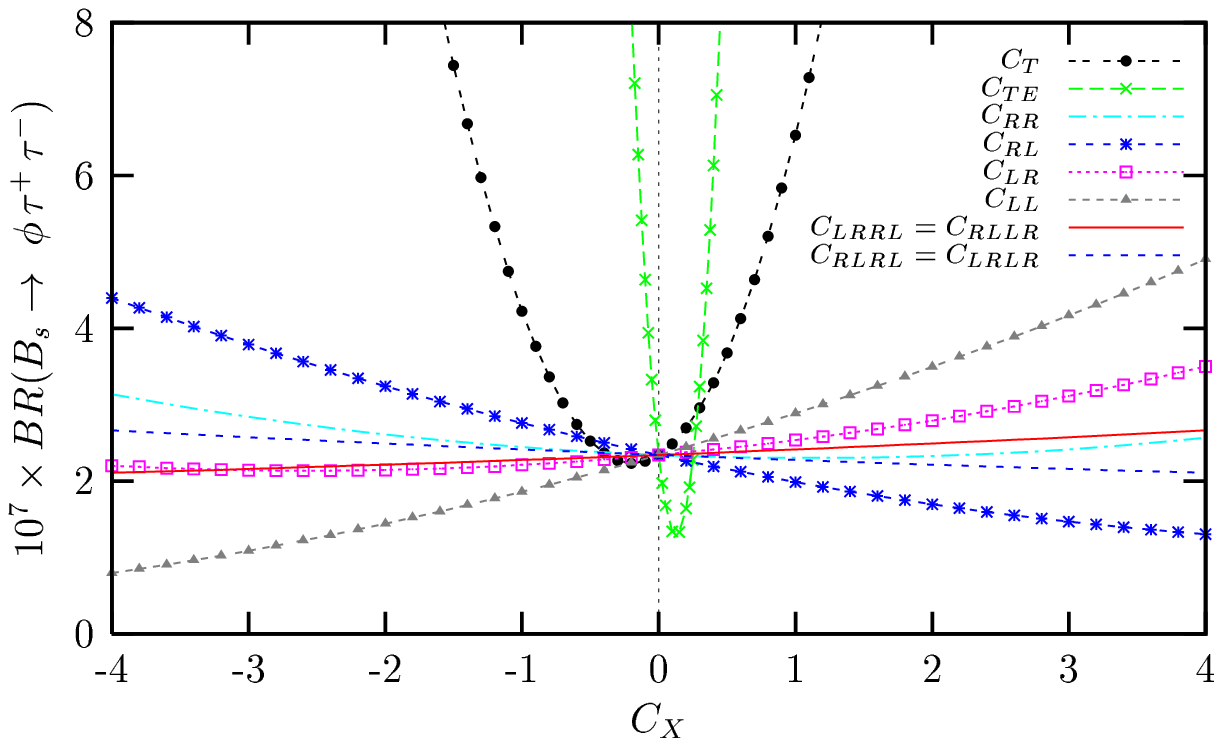}
\caption{The dependence of the integrated branching ratio for the
$B_s \rar \phi \, \tau^+ \tau^-$ decay on the new Wilson
coefficients. \label{f2}}
\end{figure}

In Figures (\ref{f1}) and (\ref{f2}), we give the dependence of the
integrated branching ratio (BR) on the new Wilson coefficients for
the \Bphimu and \Bphitau decays, respectively.
The SM predictions for the integrated branching ratios, which are comparable
with \cite{{Deandrea}, {Gursevil2002}},  are
\bea
BR (B_s \rar \phi\, \mu^+ \mu^-)   & = & 1.92 \times 10^{-6}\, , \nnb \\
BR (B_s \rar \phi\, \tau^+ \tau^-) & = & 2.34 \times 10^{-7}\, .\nnb
\eea
The former one is also in the experimental range reported by \cite{DO2006}-\cite{CDF2008}.

In figures, the strong dependence of BR on the tensor interactions
is clear. There is a weak dependence on vector interactions $C_{LL}$
and $C_{RL}$, while BR is completely insensitive to the scalar
interactions for $\mu$ case, negligibly sensitive for $\tau$ case.
It can also be seen from these figures that dependence of the BR on
the new Wilson coefficients is symmetric with respect to the zero
point for the muon final state, but such a symmetry is not observed
for the tau final state for the tensor interactions.

\begin{figure}
\centering
\includegraphics[width=5in]{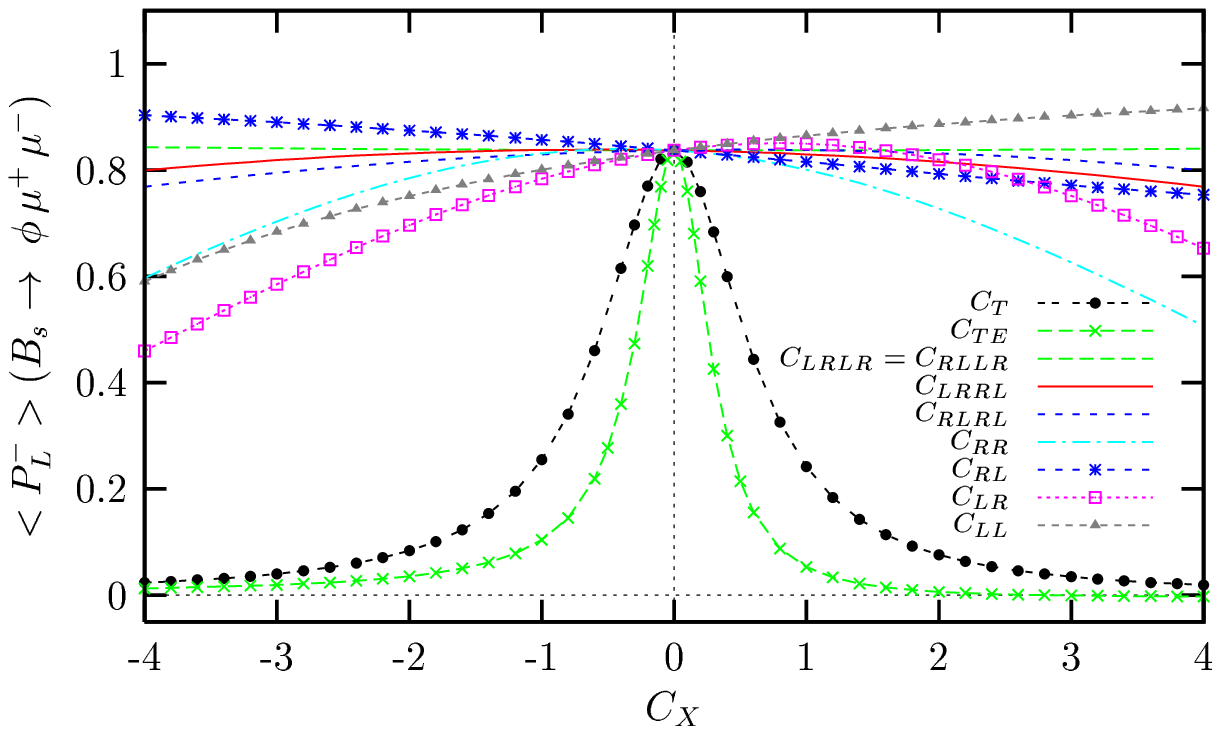}
\caption{The dependence of the averaged longitudinal polarization
$<P^-_L>$ of $\ell^-$ for the $B_s \rar \phi \, \mu^+ \mu^-$ decay
on the new Wilson coefficients. \label{f3}}
\end{figure}

\begin{figure}
\centering
\includegraphics[width=5in]{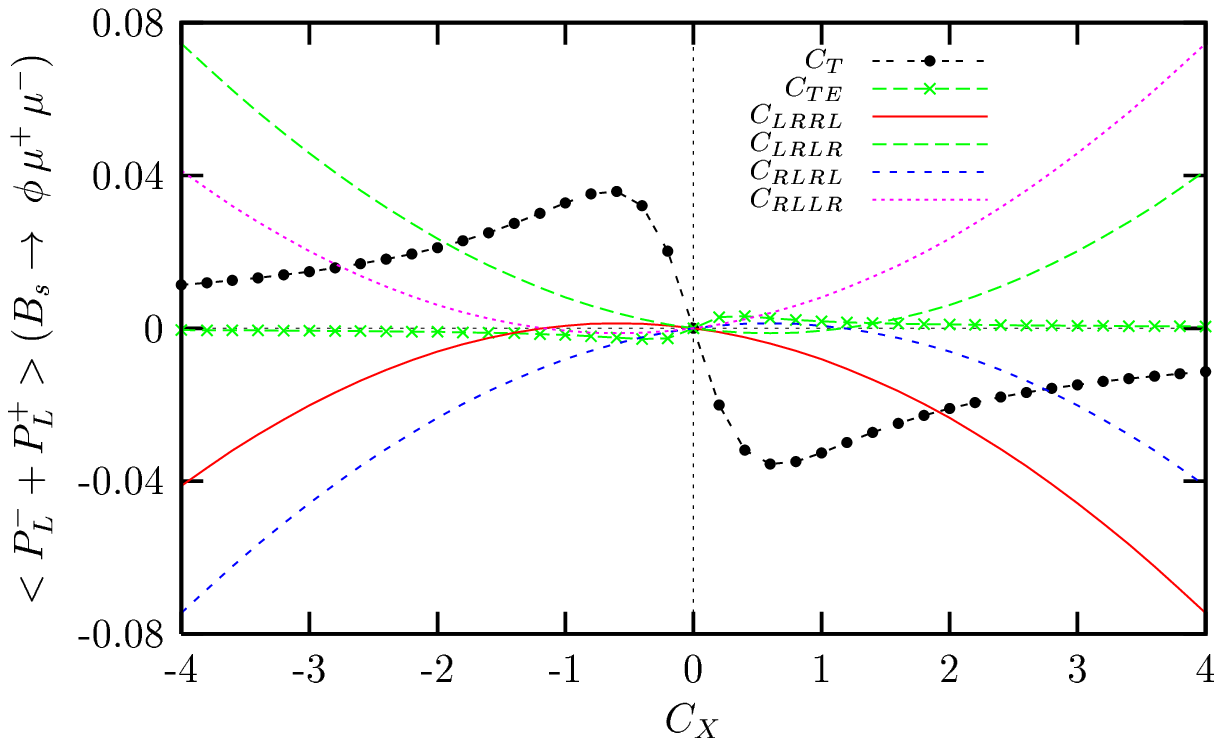}
\caption{The dependence of the combined averaged longitudinal lepton
polarization \, \, \, $<P^-_L+P^+_L>$ for the $B_s \rar \phi \,
\mu^+ \mu^-$  decay on the new Wilson coefficients.\label{f4}}
\end{figure}

In Figs. (\ref{f3}) and (\ref{f4}), we present the dependence of
averaged longitudinal polarization $<P_L^->$ of $\ell^-$ and the
combined averaged $<P_L^- + P_L^+ >$ for \Bphimu decay on the new
Wilson coefficients. We observe that the contributions
coming from all types of interactions to $<P_L^- >$ are positive, and more sensitive to the
existence of the tensor interactions. It is an increasing (decreasing)
functions of both tensor interactions for their negative (positive) values and it should also be noted that
$<P_L^- >$ becomes substantially
different from the SM value (at $C_X=0$) as $C_X$ becomes different
from zero. This indicates that measurement of the longitudinal
lepton polarization in  \Bphimu decay can
be very useful to investigate new physics beyond the SM.
On the other
hand, the contributions to the combined average $<P_L^- + P_L^+ >$
is a result of scalar and $C_T$ tensor interactions. Vector type interactions are cancelled when the
longitudinal polarization asymmetry of the lepton and antilepton are
considered together. This expected
result is also tested here since there is no vector
contribution on $<P_L^- + P_L^+>$.
Additionally, $<P_L^- + P_L^+ >$ becomes zero at $C_X=0$, which conforms the SM
results. So, any nonzero measurement  can be the signal of new physics beyond the SM.
The dependence of $<P_L^- + P_L^+ >$ on $C_X$ is symmetric with respect to
the zero value and  is positive for all values of $C_{LRLR}$ and $C_{RLLR}$,
while it is negative for remaining scalar type interactions. A last note on the $C_T$ interaction.
The $C_T$ contribution is positive (negative) for $C_X<0$ ($C_X>0$). This can also be useful.

\begin{figure}
\centering
\includegraphics[width=5in]{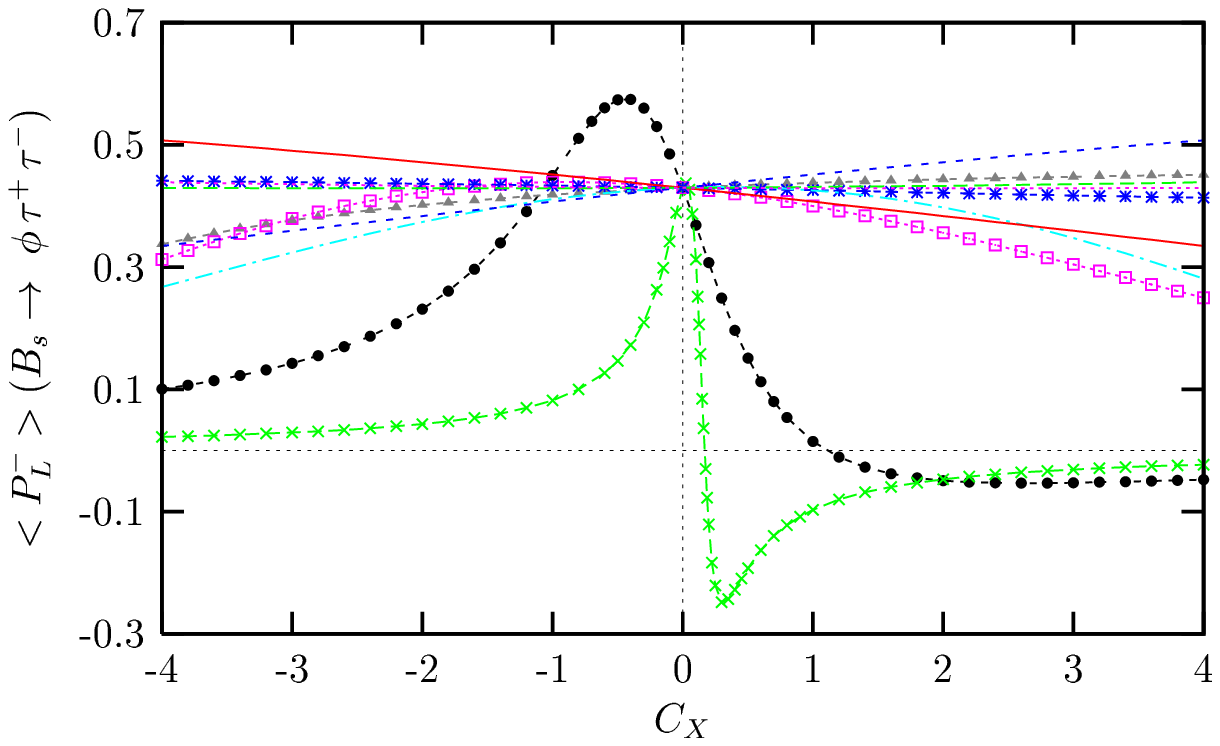}
\caption{The same as Fig. (\ref{f3}), but for the $B_s \rar
 \phi \, \tau^+ \tau^-$ decay. \label{f5}}
\end{figure}

\begin{figure}
\centering
\includegraphics[width=5in]{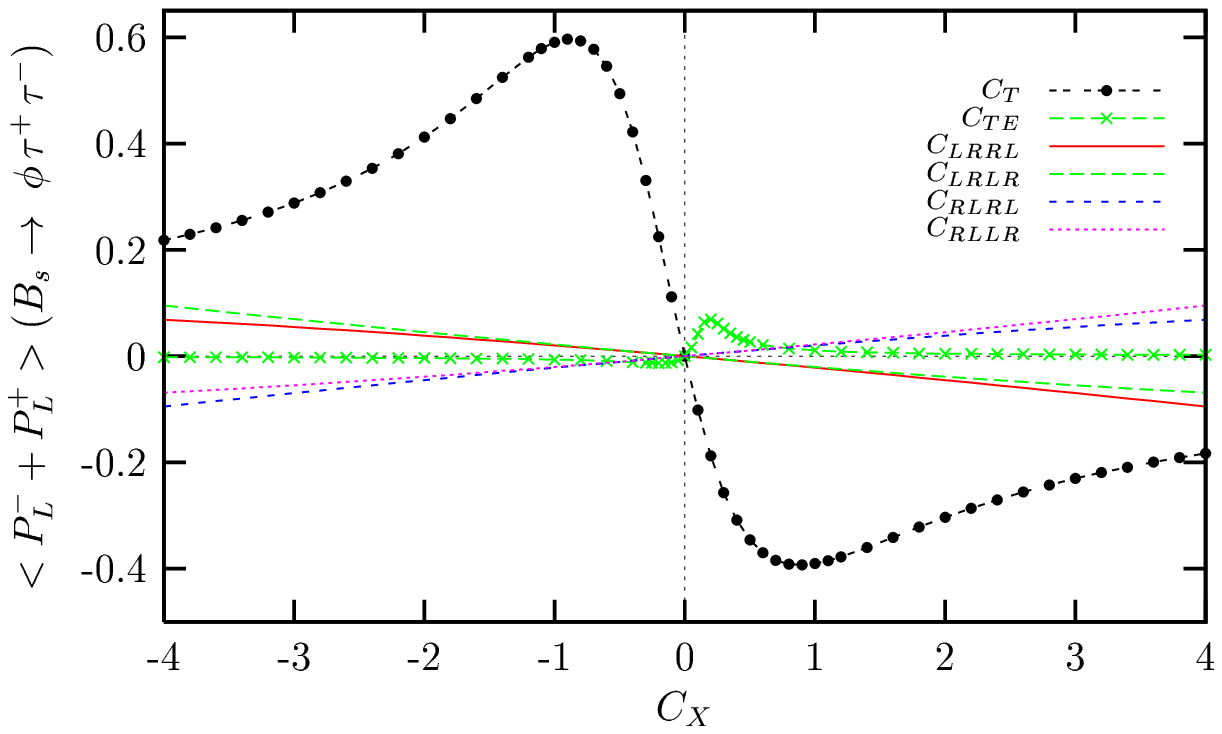}
\caption{The same as Fig. (\ref{f4}), but for the $B_s \rar
 \phi \, \tau^+ \tau^-$ decay. \label{f6}}
\end{figure}

Figures (\ref{f5}) and (\ref{f6}) are the same as Figs. (\ref{f3})
and (\ref{f4}), but for \Bphitau.
Similar to
the muon case, $<P_L^- >$ is more  sensitive to the tensor
interactions than others.
Contributions to $<P_L^- >$  from all type of interactions are positive
for all values of $C_X$ except for $C_T$ and $C_{TE}$. In the region
$1.2\simlt C_T <4$ and $0.17\simlt C_{TE}<4$, $<P_L^- >$ changes the
sign and becomes negative.
For $<P_L^- + P_L^+ >$, although their values are bigger than that of the $\mu$ case, the scaler and $C_{TE}$ tensor
contributions become less important as comparing the dominance of $C_{TE}$.
$<P_L^- + P_L^+ >$ changes sign as individual Wilson coefficient changes its sign. Specifically
speaking, the $<P_L^- + P_L^+ >$ takes positive (negative) values for negative (positive) value of $C_{T}$.
Thus, one can provide essential
information about new physics by determining the sign and the magnitude of $<P_L^- + P_L^+ >$.
In tau final state, $<P_L^- + P_L^+ >$ also becomes
zero at $C_X=0$, a conforming result of the SM.

\begin{figure}
\centering
\includegraphics[width=5in]{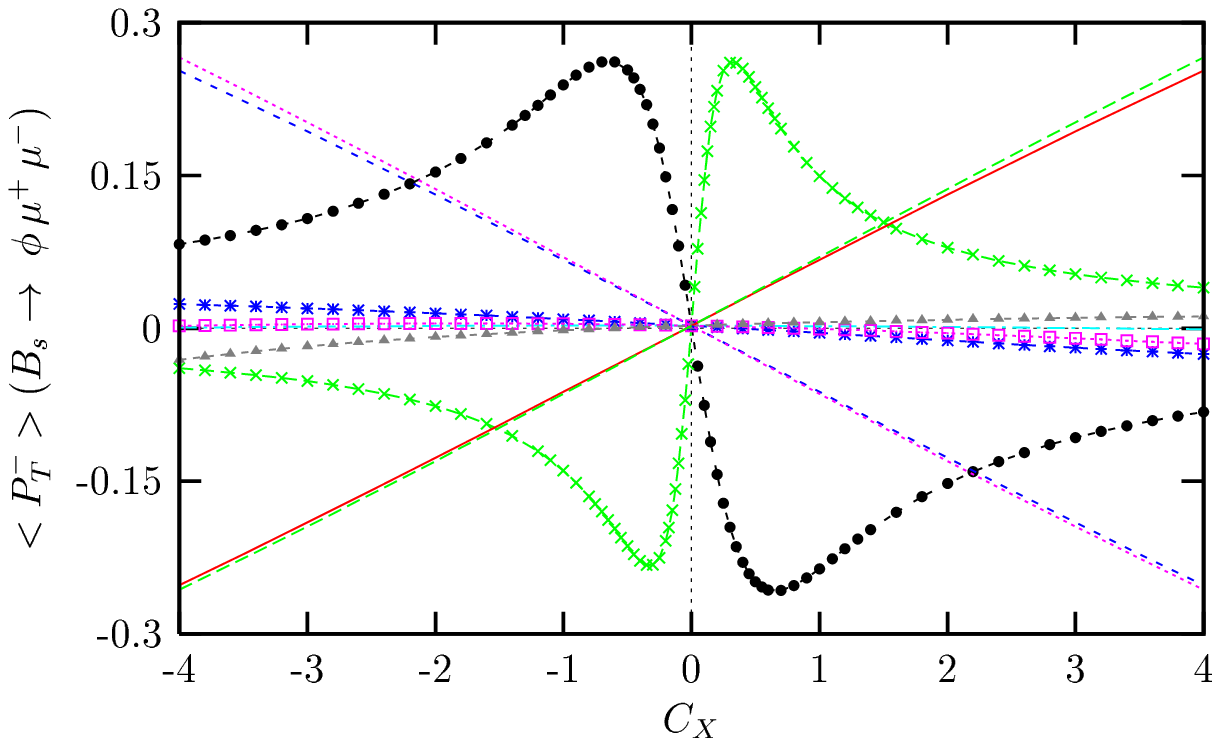}
\caption{The dependence of the averaged transverse polarization
$<P^-_T>$ of $\ell^-$ for the $B_s \rar \phi \, \mu^+ \mu^-$ decay
on the new Wilson coefficients. \label{f7}}
\end{figure}

\begin{figure}
\centering
\includegraphics[width=5in]{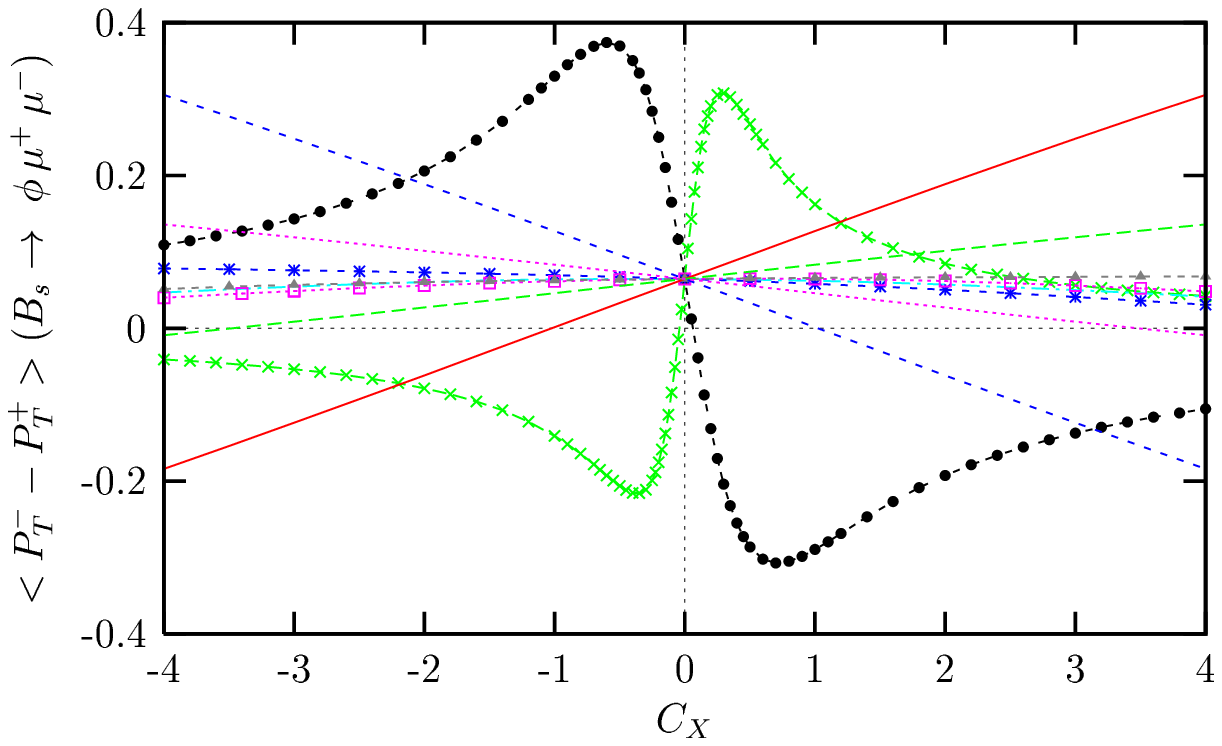}
\caption{The dependence of the combined averaged transverse lepton
polarization \, \, \, $<P^-_T-P^+_T>$ for the $B_s \rar \phi\, \mu^+ \mu^-$  decay on the new Wilson coefficients.\label{f8}}
\end{figure}

In Figs. (\ref{f7}) and (\ref{f8}), we present the dependence of
averaged transverse polarization $<P_T^->$ of $\ell^-$ and the
combined averaged $<P_T^- - P_T^+ >$ for \Bphimu decay on the new
Wilson coefficients. As seen from the figure, for  $<P_T^- >$ the vector contributions are
negligible but there appears strong dependence
on tensor and scaler interactions. The scalar terms $C_{LRRL}$, $C_{LRLR}$ and $C_{RLRL}$, $C_{RLLR}$
are approximately identical in pair. When the formers are positive (negative), $<P_T^- >$ is negative (positive)
while it is opposite for the others.
On the other hand, the $C_{LRLR}$ and $C_{RLRL}$
components of scaler interactions become less important in $<P_T^- - P_T^+ >$, as comparing their effects in $<P_T^- >$.

\begin{figure}
\centering
\includegraphics[width=5in]{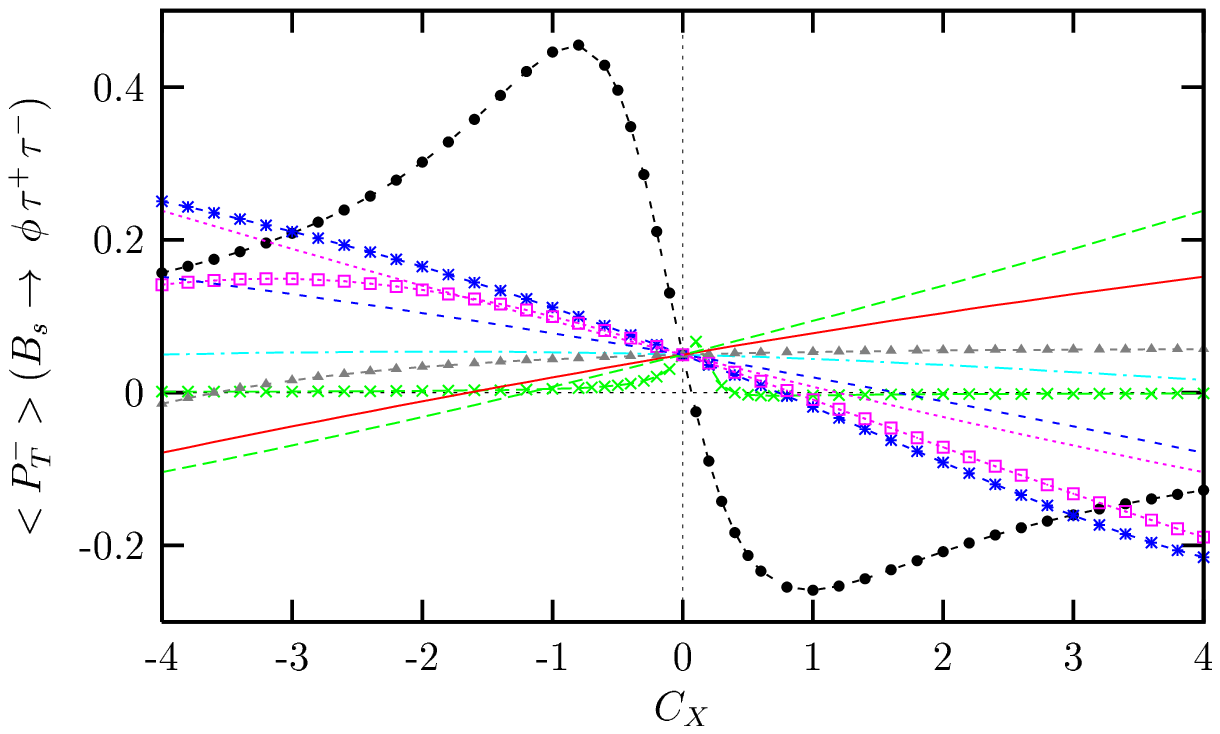}
\caption{The same as Fig. (\ref{f7}), but for the $B_s \rar
 \phi \, \tau^+ \tau^-$ decay. \label{f9}}
\end{figure}

\begin{figure}
\centering
\includegraphics[width=5in]{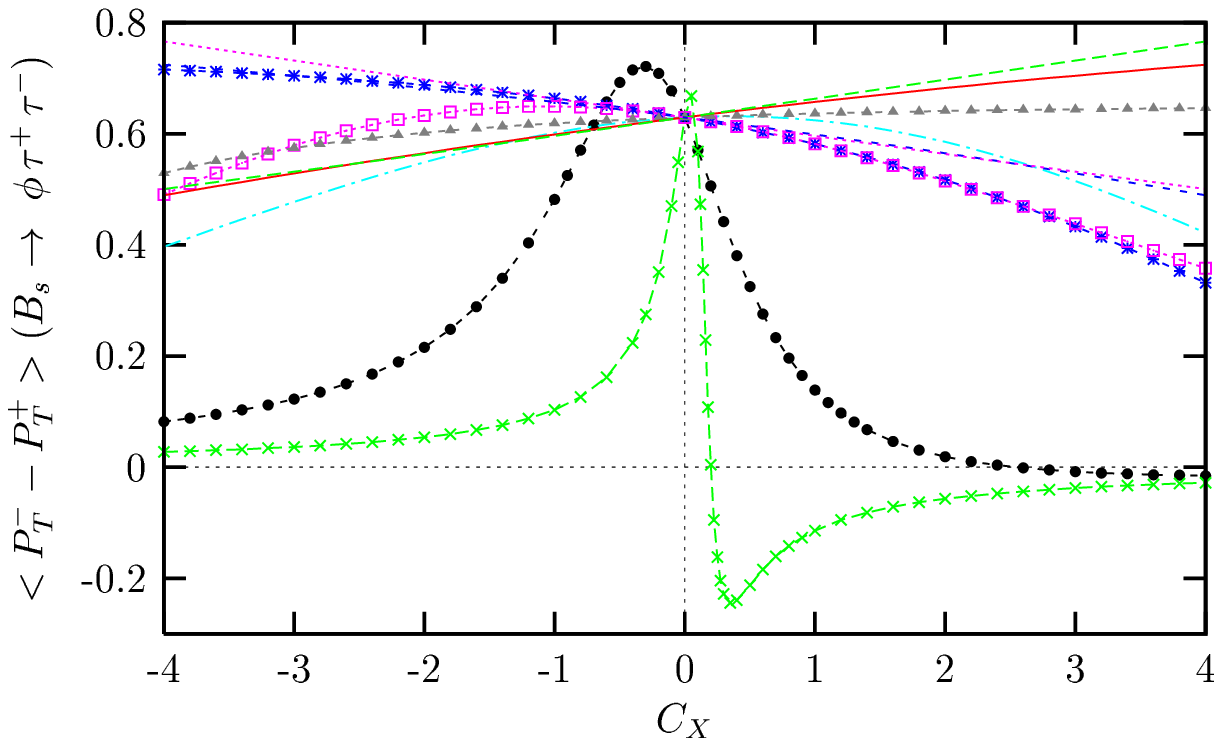}
\caption{The same as Fig. (\ref{f8}), but for the $B_s \rar
 \phi \, \tau^+ \tau^-$ decay.\label{f10}}
\end{figure}

Figures (\ref{f9}) and (\ref{f10}) are the same as Figs. (\ref{f7})
and (\ref{f8}), but for \Bphitau. We see from
these figures that the $<P_T^->$ and $<P_T^- - P_T^+>$ are quite sensitive to all types of interactions.
The dependence of vector interactions are also more sizable comparing with the muon final state case.
Changes in sign of $<P_T^->$ and $<P_T^- - P_T^+>$ are observed depending on the change in the Wilson
coefficients, the measurement of which may be useful to search new physics.

\begin{figure}
\centering
\includegraphics[width=5in]{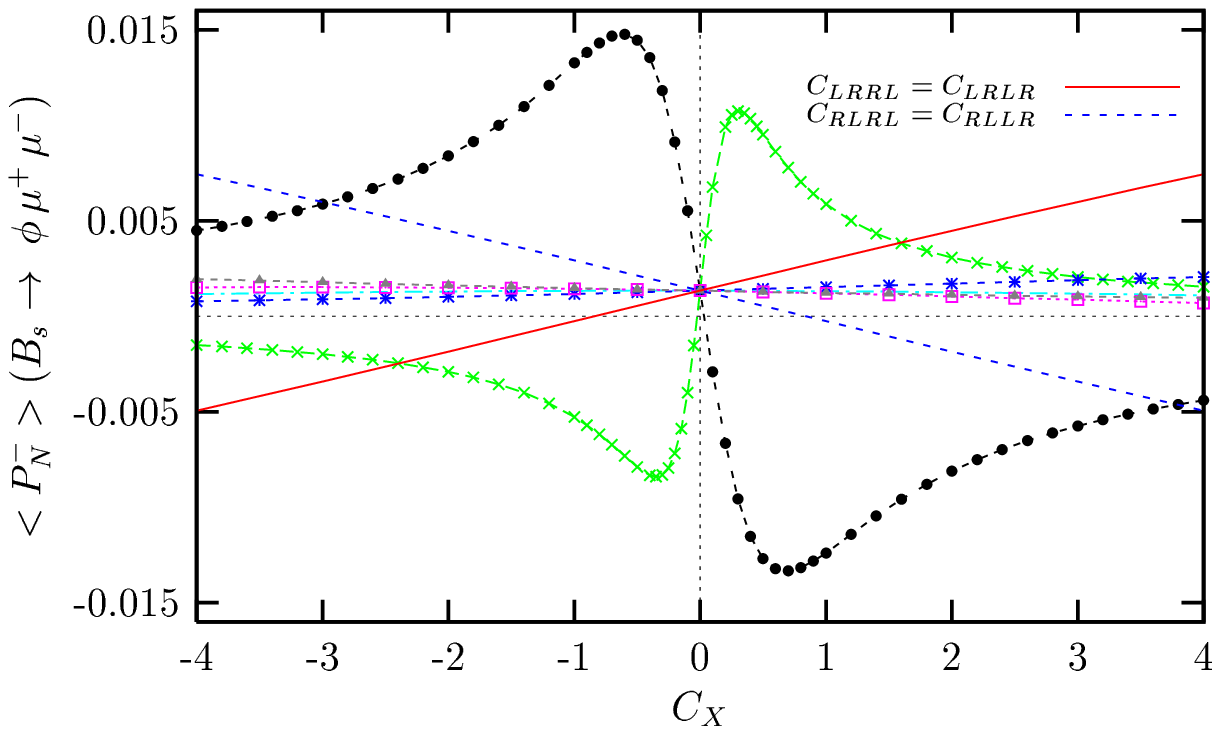}
\caption{The dependence of the averaged normal polarization
$<P^-_N>$ of $\ell^-$ for the $B_s \rar \phi \, \mu^+ \mu^-$ decay
on the new Wilson coefficients.\label{f11}}
\end{figure}

\begin{figure}
\centering
\includegraphics[width=5in]{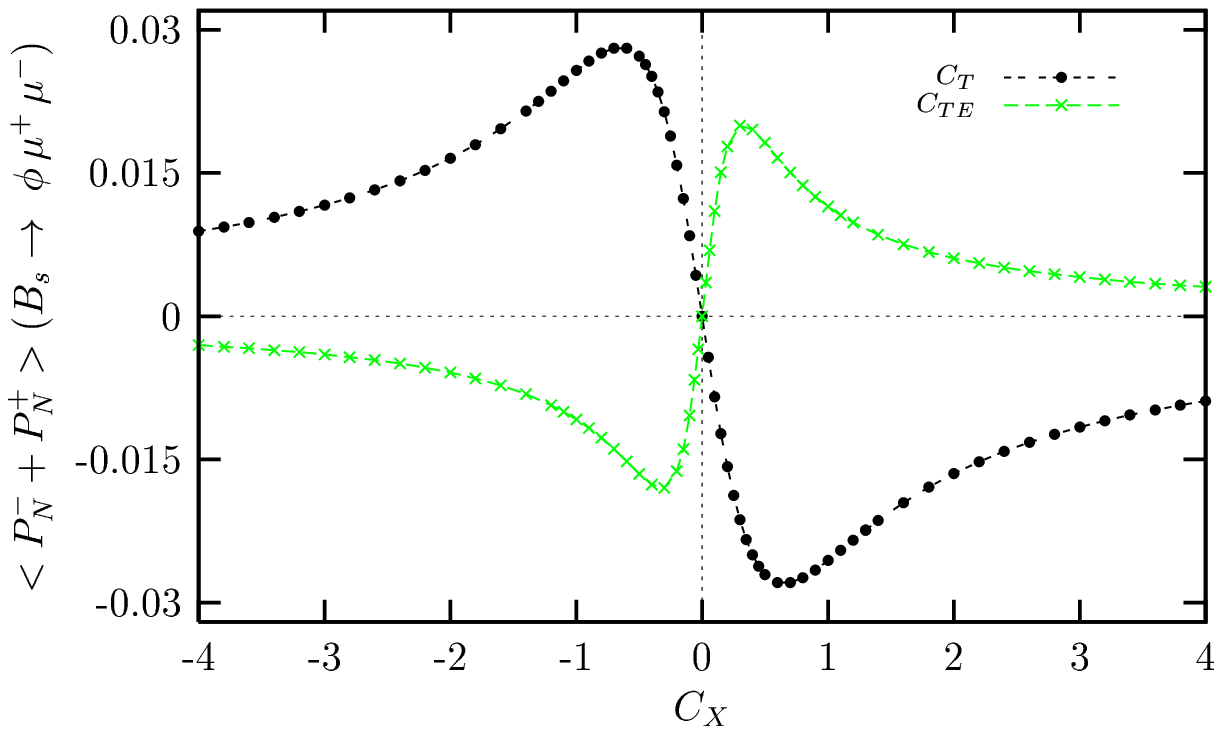}
\caption{The dependence of the combined averaged normal lepton
polarization $<P^-_N+P^+_N>$ for the $B_s \rar \phi \, \mu^+
\mu^-$  decay on the new Wilson coefficients.\label{f12}}
\end{figure}

In Figs. (\ref{f11}) and (\ref{f12}), we present the dependence of
averaged normal polarization $<P_N^->$ of $\ell^-$ and the combined
averaged $<P_N^- + P_N^+>$ for \Bphimu decay on the new Wilson
coefficients.
We see from Fig.  (\ref{f11}) that
$<P_N^->$ strongly depends on the tensor  and scaler interactions, the vector contribution is negligible.
The coefficients $C_{LRRL}, C_{LRLR}$ and $C_{RLLR}, C_{RLRL}$ are identical in pairs.
the behavior of $<P_N^- + P_N^+ >$ is determined by  the tensor
interactions only. We observe that $<P_N^- + P_N^+ >$ is negative (positive)  when $C_{TE}<0$ ($C_{TE}>0$)
while  the behavior of $C_{T}$ is opposite with respect to $C_{TE}$. In addition, as expected in the SM,
$<P_N^- + P_N^+>$ becomes zero at $C_X=0$.

\begin{figure}
\centering
\includegraphics[width=5in]{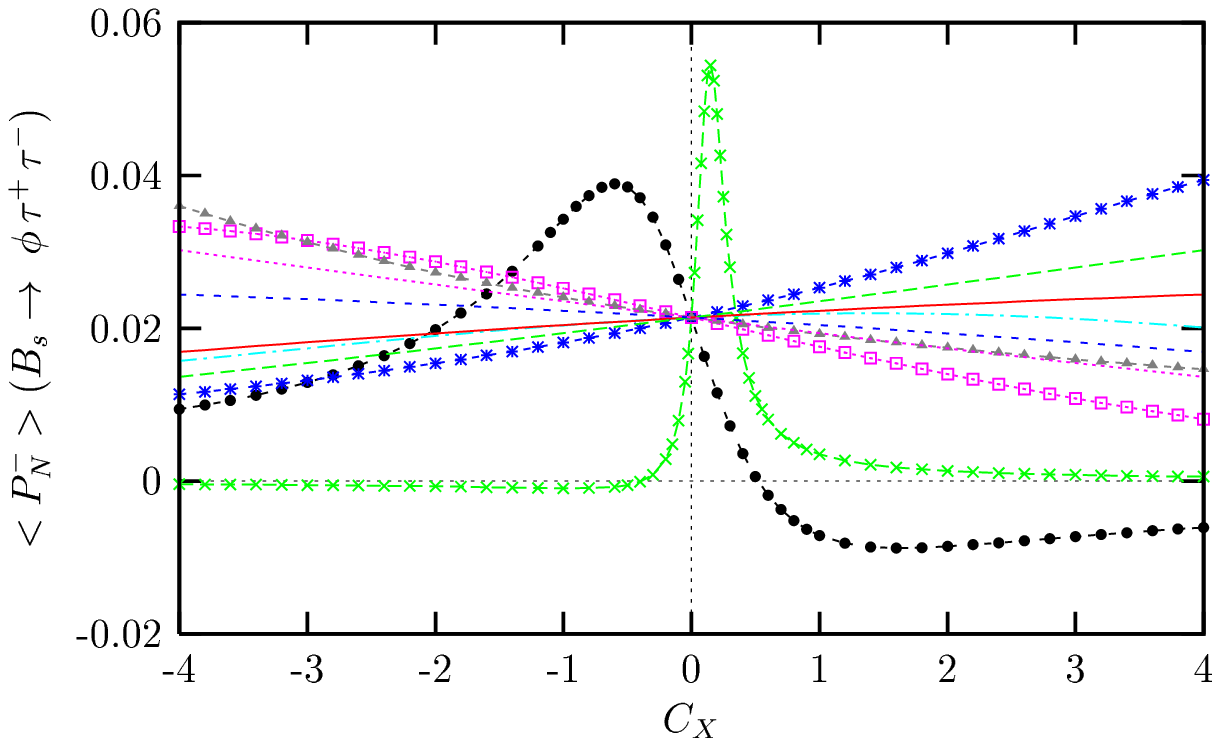}
\caption{The same as Fig.(\ref{f11}), but for the $B_s \rar \phi \,
\tau^+ \tau^-$  decay.\label{f13}}
\end{figure}

\begin{figure}
\centering
\includegraphics[width=5in]{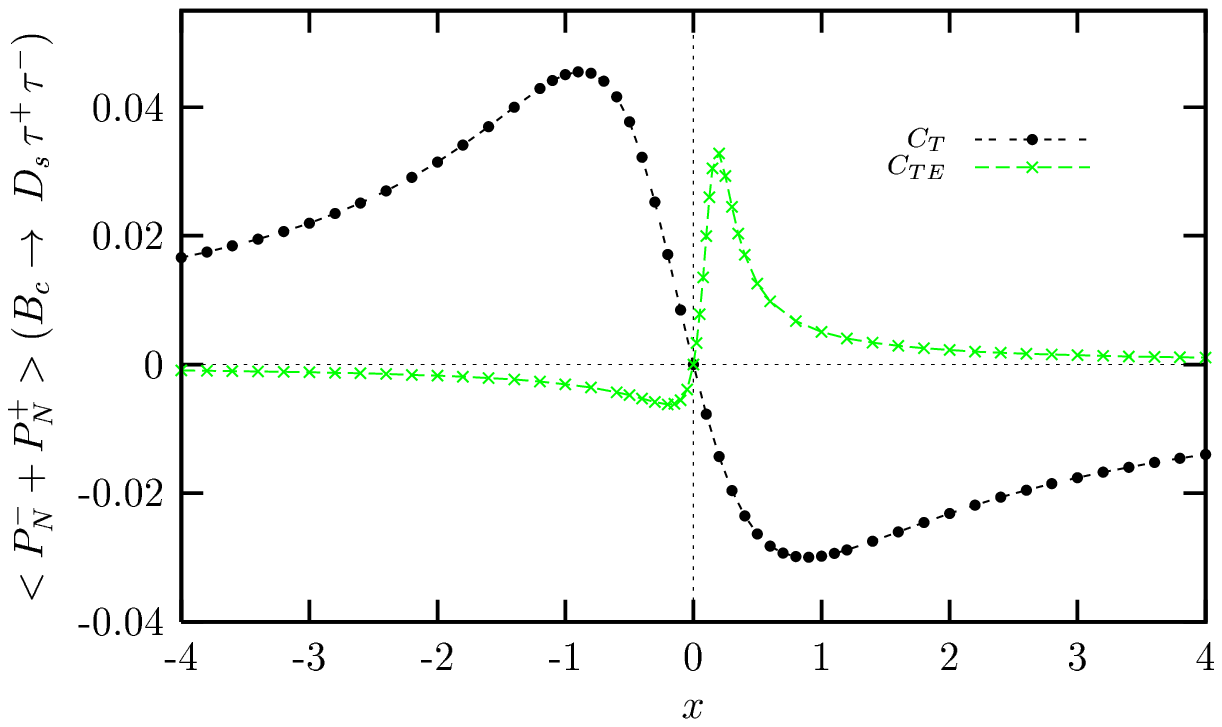}
\caption{The same as Fig. (\ref{f12}), but for the $B_s \rar
 \phi \, \tau^+ \tau^-$ decay.\label{f14}}
\end{figure}

Figures (\ref{f13}) and (\ref{f14}) are the same as Figs. (\ref{f11})
and (\ref{f12}), but for \Bphitau.
In opposite to the muon final state case, we should notice the dependence of $<P_N^->$ on vector type interactions, too.
$<P_N^- + P_N^+>$, as in the muon case, depends only on the tensor
interactions and their behaviors are similar.

Finally, a few words on the detectibility of lepton polarization
asymmetries to have an idea of this possibility folowing \cite{Geng96}.
Experimentally, the required number of events are $N=n^2/(B<P_i>^2$
for a decay with the branching ration B at $n\sigma$ level to be able to measure an asymmery $<P_i>$.
Using our SM predictions for lepton polarizations given in (\ref {smpre}) we simply find that to observe
$<P_L^->, <P_T^->$ and $<P_N^->$ in \Bphimu at $1\sigma$ level we need $N=(0.74;8.33\times10^4;3.08\times10^5)\times10^6$
number of events, respectively. For the \Bphitau, the required number of
events are $N=(2.31;1.72\times10^2;9.33\times10^2)\times10^7$.
The number of $b\bar b$ events expected, at least at LHC-b, is $\sim 10^{12}$. So, comparing these numbers
we conclude that in principle measurement of these values could be possible.

In conclusion, starting the general model independent form of the
effective Hamiltonian, we present the most general analysis of the
lepton polarization asymmetries in the rare \Bphi decay. The
dependence of the longitudinal, transversal and normal polarization
asymmetries of $\ell^-$ and their combined asymmetries on the new
Wilson coefficients are studied. The lepton polarization asymmetries
are very sensitive to the existence of the tensor type interactions
and in some cases effect of scaler type interactions should be considered.
The tensor $C_{T}$ term plays a significant role
throughout this work. Additionally, in the most
cases, the value of polarization asymmetries change sign  as the new Wilson
coefficients vary in the region of interest, which is useful to
determine the sign in addition magnitude of new physics effect.
In the SM, in the limit $m_\ell \rar 0$, the combined lepton polarizations are
$<P_L^- + P_L^+> =0$, $<P_N^- + P_N^+>=0$ and $<P_T^- - P_T^+>\simeq 0$.
Therefore, in the experimental searches, any nonzero measurement will be an
effective tool in looking for new physics beyond the SM.
\\
\\
{\large \bf Acknowledgments}\\
The author would like to thank S. Fajfer for valuable contributions and critical comments,
G. Turan for reading the manuscript and comments on it and
B. Golob for sharing experimental experience.
This work was supported by The Scientific and
Technological Research Council of Turkey (TUBITAK) under BIDEB-2219 program.
\\
\newpage
\renewcommand{\theequation}{A-\arabic{equation}}
  \setcounter{equation}{0}  
  \section*{Appendix A}  
\bea  \label{delta} \Delta &=&
    \frac{32}{3} m_{B_s}^4 \lambda
    \Big[(m_{B_s}^2 s -
    m_\ell^2) \ga \vel  A_1 \ver^2 + \vel  C_1 \ver^2 \dr + 6 m_\ell^2
    \, \mbox{\rm Re}
    (A_1 C_1^\ast)\Big]  \nnb \\
&+& 96 m_\ell^2 \, \mbox{\rm Re} (B_1 D_1^\ast)-
    \frac{4}{r} m_{B_s}^2 m_\ell \lambda \,
    \mbox{\rm Re} [(B_1 - D_1) (B_4^\ast - B_5^\ast)] \nnb \\
&+&\frac{8}{r} m_{B_s}^2 m_\ell^2 \lambda \, \Big(
    \mbox{\rm Re} [B_1 (- B_3^\ast + D_2^\ast + D_3^\ast)] +
    \mbox{\rm Re} [D_1 (B_2^\ast + B_3^\ast - D_3^\ast)] -
    \mbox{\rm Re}(B_4 B_5^\ast) \Big)~~~~~~~~ \nnb \\
&+&\frac{4}{r} m_{B_s}^4 m_\ell (1-r) \lambda \,
    \Big(\mbox{\rm Re} [(B_2 - D_2) (B_4^\ast - B_5^\ast)]
    +2 m_\ell \, \mbox{\rm Re} [(B_2 - D_2) (B_3^\ast - D_3^\ast)]
    \Big) \nnb \\
&-& \frac{8}{r}m_{B_s}^4 m_\ell^2 \lambda (2+2 r-s)\, \mbox{\rm Re}
(B_2 D_2^\ast)
    +\frac{4}{r} m_{B_s}^4 m_\ell s \lambda \,
    \mbox{\rm Re} [(B_3 - D_3) (B_4^\ast - B_5^\ast)] \nnb \\
&+&\frac{4}{r} m_{B_s}^4 m_\ell^2 s \lambda \,
    \vel B_3 - D_3\ver^2
    +\frac{2}{r} m_{B_s}^2 (m_{B_s}^2 s-2 m_\ell^2) \lambda \,
    \ga \vel B_4 \ver^2 + \vel B_5 \ver^2 \dr \nnb \\
&-&\frac{8}{3rs} m_{B_s}^2 \lambda \,
    \Big[m_\ell^2 (2-2 r+s)+m_{B_s}^2 s (1-r-s) \Big]
    \Big[\mbox{\rm Re}(B_1 B_2^\ast) + \mbox{\rm Re}(D_1 D_2^\ast)\Big] \nnb \\
&+&\frac{4}{3rs}\,
    \Big[2 m_\ell^2 (\lambda-6 rs)+m_{B_s}^2 s (\lambda+12 rs) \Big]
    \ga \vel B_1 \ver^2 + \vel D_1 \ver^2 \dr \nnb \\
&+&\frac{4}{3rs} m_{B_s}^4 \lambda\,
    \Big( m_{B_s}^2 s \lambda + m_\ell^2 [ 2 \lambda + 3 s (2+2 r - s) ] \Big)
    \ga \vel B_2 \ver^2 + \vel D_2 \ver^2 \dr \nnb \\
&+&\frac{32}{r} m_{B_s}^6 m_\ell \lambda^2 \,
    \mbox{\rm Re} [(B_2 + D_2)(B_7 C_{TE})^\ast]  \nnb \\
&-& \frac{32}{r} m_{B_s}^4 m_\ell \lambda (1-r-s) \Big(
    \mbox{\rm Re} [(B_1 + D_1)(B_7 C_{TE})^\ast] +
    2\, \mbox{\rm Re} [(B_2 + D_2)(B_6 C_{TE})^\ast] \Big) \nnb \\
&+&\frac{64}{r} (\lambda+12 rs) m_{B_s}^2 m_\ell \,
    \mbox{\rm Re} [(B_1 + D_1)(B_6 C_{TE})^\ast] \nnb \\
&+&\frac{256}{3rs} \vel T_1 \ver^2 \vel C_T \ver^2 m_{B_s}^2
    \Big( 4 m_\ell^2 \, [ \lambda ( 8r -s) - 12 r s (2 +2r -s) ] \nnb \\
&+& m_{B_s}^2 s \, [\lambda (16r+s)+12 r s (2 +2r -s) ] \Big) \nnb \\
&+&\frac{1024}{3rs} \vel T_1 \ver^2 \vel C_{TE} \ver^2 m_{B_s}^2
    \Big( 8 m_\ell^2 \, [ \lambda ( 4r+s) + 12 r s (2 +2r -s) ] \nnb \\
&+& m_{B_s}^2 s \, [\lambda (16r+s)+12 r s (2 +2r -s) ] \Big) \nnb \\
&-& \frac{128}{r} m_{B_s}^2 m_\ell \, [\lambda + 12 r (1-r) ]
    \,\mbox{\rm Re} [(B_1 + D_1)(T_1 C_{TE})^\ast] \nnb \\
&+&\frac{128}{r} m_{B_s}^4 m_\ell \lambda (1+3r-s)
    \,\mbox{\rm Re} [(B_2 + D_2)(T_1 C_{TE})^\ast]
    + 512 m_{B_s}^4 m_\ell \lambda \,
    \mbox{\rm Re} [(A_1 + C_1)(T_1 C_T)^\ast] \nnb \\
&+&\frac{16}{3r} m_{B_s}^2 \Bigg( 4 (m_{B_s}^2 s + 8 m_\ell^2) \vel
C_{TE} \ver^2
    + m_{B_s}^2 s v^2 \vel C_T \ver^2 \Bigg)
    \times \Bigg( 4 (\lambda+12 r s) \vel B_6 \ver^2 \nnb \\
&+& m_{B_s}^4 \lambda^2 \vel B_7 \ver^2
    - 4 m_{B_s}^2 (1-r-s) \lambda \,\mbox{\rm Re} (B_6 B_7^\ast)
    - 16 \, [\lambda + 12 r (1-r) ]\,\mbox{\rm Re} (T_1 B_6^\ast) \nnb \\
&+& 8 m_{B_s}^2 (1+3r-s) \lambda \,\mbox{\rm Re} (T_1
    B_7^\ast)\Bigg)~,
\eea
where $\lambda= 1 + r^2 + s^2 -2r-2s-2rs$, $r=m_{\phi}^2/m_{B_s}^2$
lepton velocity is $v=\sqrt{1-{4m_\ell^2}/{s m_{B_s}^2}}$.

\newpage
\renewcommand{\theequation}{B-\arabic{equation}}
  \setcounter{equation}{0}  
  \section*{Appendix B}  
The longitudinal polarization $P_L^{\pm}$ for the $\ell^{\pm}$;
\bea  \label{plm} P_L^{\pm}&=& \frac{4}{\Delta} m_{B_s}^2 v
    \Bigg\{\mp
    \frac{1}{3 r} \lambda^2 m_{B_s}^4 \Big[ \vel B_2 \ver^2 - \vel D_2
    \ver^2\Big] + \frac{1}{r} \lambda  m_\ell \,
    \mbox{\rm Re} [(B_1 - D_1) (B_4^\ast + B_5^\ast)] \nnb \\
&-& \frac{1}{r} \lambda m_{B_s}^2 m_\ell (1-r) \,
    \mbox{\rm Re} [(B_2 - D_2) (B_4^\ast + B_5^\ast)]\mp
    \frac{8}{3} \lambda m_{B_s}^4 s \Big[ \vel A_1 \ver^2 - \vel C_1 \ver^2\Big]
    \nnb \\
&-&\frac{1}{2r} \lambda m_{B_s}^2 s
    \Big[ \vel B_4 \ver^2 - \vel B_5 \ver^2\Big]-
    \frac{1}{r} \lambda m_{B_s}^2 m_\ell s \,
    \mbox{\rm Re} [(B_3 - D_3) (B_4^\ast + B_5^\ast)] \nnb \\
&\pm &\frac{2}{3 r} \lambda m_{B_s}^2 (1-r-s)
    \Big[ \mbox{\rm Re}(B_1 B_2^\ast) - \mbox{\rm Re}(D_1 D_2^\ast)\Big]
    \mp\frac{1}{3 r} (\lambda + 12 r s)
    \Big[ \vel B_1 \ver^2 - \vel D_1 \ver^2\Big] \nnb \\
&\mp&\frac{256}{3} \lambda m_{B_s}^2 m_\ell
    \,\Big( \mbox{\rm Re} [A_1^\ast (C_{T} \mp C_{TE})T_1] -
    \mbox{\rm Re} [C_1^\ast (C_T\pm C_{TE})T_1] \Big) \nnb \\
&+&\frac{4}{3r} \lambda^2 m_{B_s}^4 m_\ell \Big(
    \mbox{\rm Re} [B_2^\ast (C_T\mp 4 C_{TE}) B_7]
    + \mbox{\rm Re} [D_2^\ast (C_T\pm 4 C_{TE}) B_7] \Big)  \nnb \\
&-&\frac{8}{3r} \lambda m_{B_s}^2 m_\ell (1-r-s) \Big(
    \mbox{\rm Re} [B_2^\ast (C_T\mp 4 C_{TE}) B_6]
    +\mbox{\rm Re} [D_2^\ast (C_T\pm 4 C_{TE}) B_6] \Big) \nnb \\
&-&\frac{4}{3r} \lambda m_{B_s}^2 m_\ell (1-r-s) \Big(
    \mbox{\rm Re} [B_1^\ast (C_T\mp 4 C_{TE}) B_7]
    + \mbox{\rm Re} [D_1^\ast (C_T\pm 4 C_{TE}) B_7]  \Big) \nnb \\
&+&\frac{8}{3r} (\lambda+12 r s)  m_\ell \Big(
    \,\mbox{\rm Re} [B_1^\ast (C_T\mp 4 C_{TE}) B_6]
    +\mbox{\rm Re} [D_1^\ast (C_T\pm 4 C_{TE}) B_6] \Big) \nnb \\
&-&\frac{16}{3r}  m_\ell [\lambda+12 r (1-r)] \Big(
    \mbox{\rm Re} [B_1^\ast (C_T\mp 4 C_{TE}) T_1]
    +\mbox{\rm Re} [D_1^\ast (C_T\pm 4 C_{TE}) T_1] \Big) \nnb \\
&+&\frac{16}{3r} \lambda m_{B_s}^2 m_\ell (1+3 r - s) \Big(
    \mbox{\rm Re} [B_2^\ast (C_T\mp 4 C_{TE}) T_1]
    +\mbox{\rm Re} [D_2^\ast (C_T\pm 4 C_{TE}) T_1] \Big) \nnb \\
&+&\frac{16}{3r} \lambda^2 m_{B_s}^6 s
    \vel B_7 \ver^2 \mbox{\rm Re} (C_T C_{TE}^\ast)
  +\frac{64}{3r} (\lambda + 12 r s) m_{B_s}^2 s
    \vel B_6 \ver^2 \mbox{\rm Re} (C_T C_{TE}^\ast) \nnb \\
&-&\frac{64}{3r} \lambda m_{B_s}^4 s (1-r-s)
    \,\mbox{\rm Re} (B_6 B_7^\ast) \mbox{\rm Re} (C_T C_{TE}^\ast)
    \nnb \\
&+& \frac{128}{3r} \lambda m_{B_s}^4 s (1+3 r-s)
    \,\mbox{\rm Re} (B_7 T_1{^\ast)} \mbox{\rm Re} (C_T C_{TE}^\ast) \nnb \\
&-&\frac{256}{3r} m_{B_s}^2 s [\lambda + 12 r (1-r)]
    \,\mbox{\rm Re} (B_6 T_1{^\ast}) \mbox{\rm Re} (C_T C_{TE}^\ast) \nnb \\
&+&\frac{256}{3r} m_{B_s}^2
    [ \lambda (4 r + s) + 12 r (1-r)^2 ] \vel T_1 \ver^2
    \,\mbox{\rm Re} (C_T C_{TE}^\ast)
    \Bigg\}~,
\eea
where $\Delta$ is given in (\ref{delta}).

The transverse polarization $P_T^{-}$ for $\ell^-$;
\bea  \label{ptm}
    P_T^- &=& \frac{\pi}{\Delta} m_{B_s} \sqrt{s \lambda} \Bigg\{
    -8 m_{B_s}^2 m_\ell  \, \mbox{\rm Re} [(A_1 + C_1) (B_1^\ast +
    D_1^\ast)] \nnb \\
 &+&  \frac{1}{r} m_{B_s}^2 m_\ell (1+3 r - s)  \,
    \Big[ \mbox{\rm Re}(B_1 D_2^\ast) -  \mbox{\rm Re}(B_2 D_1^\ast)\Big]  \nnb \\
&+&\frac{1}{r s} m_\ell (1- r - s)
    \Big[ \vel B_1 \ver^2 - \vel D_1 \ver^2\Big] \nnb \\
&+&\frac{1}{r s} (2m_\ell^2 - m_{B_s}^2 s)(1- r - s)
    \Big[ \mbox{\rm Re}(B_1 B_5^\ast) - \mbox{\rm Re}(D_1 B_4^\ast)\Big] \nnb \\
&-&\frac{1}{r} m_{B_s}^2 m_\ell (1- r - s) \,
    \mbox{\rm Re} [(B_1 + D_1) (B_3^\ast - D_3^\ast)] \nnb \\
&-&\frac{2}{r s} m_{B_s}^2 m_\ell^2 \lambda
    \Big[  \mbox{\rm Re}(B_2 B_5^\ast) -  \mbox{\rm Re}(D_2 B_4^\ast)\Big] \nnb \\
&+&\frac{1}{r s} m_{B_s}^4 m_\ell(1-r) \lambda
    \Big[ \vel B_2 \ver^2 - \vel D_2 \ver^2\Big]
    + \frac{1}{r} m_{B_s}^4 m_\ell \lambda  \,
    \mbox{\rm Re} [(B_2 + D_2) (B_3^\ast - D_3^\ast)] \nnb \\
&-&\frac{1}{r s} m_{B_s}^2 m_\ell [\lambda + (1-r-s) ( 1-r)]
    \Big[\mbox{\rm Re}(B_1 B_2^\ast) - \mbox{\rm Re}(D_1 D_2^\ast)\Big] \nnb \\
&+&\frac{1}{r s} (1-r-s)(2 m_\ell^2  - m_{B_s}^2 s )
    \Big[\mbox{\rm Re}(B_1 B_4^\ast)-\mbox{\rm Re}(D_1 B_5^\ast)\Big] \nnb \\
&+&\frac{1}{r s} m_{B_s}^2 \lambda (2 m_\ell^2  - m_{B_s}^2 s )
    \Big[\mbox{\rm Re}(D_2 B_5^\ast) - \mbox{\rm Re}(B_2 B_4^\ast)\Big] \nnb \\
&-&\frac{16}{r s} \lambda m_{B_s}^2 m_\ell^2
    \,\mbox{\rm Re}[(B_1-D_1) (B_7 C_{TE})^\ast] \nnb \\
&+&\frac{16}{r s} \lambda m_{B_s}^4 m_\ell^2 (1-r)
    \,\mbox{\rm Re}[(B_2-D_2) (B_7 C_{TE})^\ast] \nnb \\
&+&\frac{8}{r} \lambda m_{B_s}^4 m_\ell
    \,\mbox{\rm Re}[(B_4-B_5) (B_7 C_{TE})^\ast] \nnb \\
&+&\frac{16}{r} \lambda m_{B_s}^4 m_\ell^2
    \,\mbox{\rm Re}[(B_3-D_3) (B_7 C_{TE})^\ast] \nnb \\
&+&\frac{32}{r s} m_\ell^2 (1-r-s)
    \,\mbox{\rm Re}[(B_1-D_1) (B_6 C_{TE})^\ast] \nnb \\
&-&\frac{32}{r s} m_{B_s}^2 m_\ell^2 (1-r) (1-r-s)
    \,\mbox{\rm Re}[(B_2-D_2) (B_6 C_{TE})^\ast] \nnb \\
&-&\frac{16}{r} m_{B_s}^2 m_\ell  (1-r-s)
    \,\mbox{\rm Re}[(B_4-B_5) (B_6 C_{TE})^\ast] \nnb \\
&-&\frac{32}{r} m_{B_s}^2 m_\ell^2  (1-r-s)
    \,\mbox{\rm Re}[(B_3-D_3) (B_6 C_{TE})^\ast] \nnb \\
&-& 16 m_{B_s}^2  \Big(
    4 m_\ell^2 \, \mbox{\rm Re}[A_1^\ast (C_T+2 C_{TE}) B_6]
    - m_{B_s}^2 s \, \mbox{\rm Re}[A_1^\ast (C_T-2 C_{TE}) B_6] \Big)\nnb \\
&+& 16 m_{B_s}^2  \Big(
    4 m_\ell^2 \, \mbox{\rm Re}[C_1^\ast (C_T-2 C_{TE}) B_6]
    - m_{B_s}^2 s \, \mbox{\rm Re}[C_1^\ast (C_T+2 C_{TE}) B_6] \Big)\nnb \\
&+& \frac{32}{s} m_{B_s}^2 (1-r) \Big(
    4 m_\ell^2 \, \mbox{\rm Re}[A_1^\ast (C_T+2 C_{TE}) T_1]
    - m_{B_s}^2 s \, \mbox{\rm Re}[A_1^\ast (C_T-2 C_{TE}) T_1] \Big)\nnb \\
&-& \frac{32}{s} m_{B_s}^2 (1-r) \Big(
    4 m_\ell^2 \, \mbox{\rm Re}[C_1^\ast (C_T-2 C_{TE}) T_1]
    - m_{B_s}^2 s \, \mbox{\rm Re}[C_1^\ast (C_T+2 C_{TE}) T_1] \Big)\nnb \\
&+&\frac{64}{r s} m_{B_s}^2 m_\ell^2 (1-r) (1+3 r-s)
    \,\mbox{\rm Re}[(B_2- D_2) (T_1 C_{TE})^\ast] \nnb \\
&+&\frac{64}{r} m_{B_s}^2 m_\ell^2  (1+3 r-s)
    \,\mbox{\rm Re}[(B_3-D_3) (T_1 C_{TE})^\ast] \nnb \\
&+&\frac{32}{r} m_{B_s}^2 m_\ell  (1+3 r-s)
    \,\mbox{\rm Re}[(B_4-B_5) (T_1 C_{TE})^\ast] \nnb \\
&+&\frac{64}{r s} [m_{B_s}^2 r s - m_\ell^2 (1+7 r -s)]
    \,\mbox{\rm Re}[(B_1-D_1) (T_1 C_{TE})^\ast] \nnb \\
&-& \frac{32}{s} (4 m_\ell^2 + m_{B_s}^2 s)
    \,\mbox{\rm Re}[(B_1+D_1) (T_1 C_T)^\ast] \nnb \\
&-&2048  m_{B_s}^2 m_\ell
    \,\mbox{\rm Re}[(C_T T_1)(B_6 C_{TE})^\ast] \nnb \\
&+&\frac{4096}{s} m_{B_s}^2 m_\ell (1-r) \vel g \ver^2
    \, \mbox{\rm Re}(C_T C_{TE}^\ast)
    \Bigg\}~,
\eea
and $P_T^{+}$ for $\ell^+$
\bea
    \label{ptp}
    P_T^+&=& \frac{\pi}{\Delta} m_{B_s} \sqrt{s \lambda} \Bigg\{
    -8 m_{B_s}^2 m_\ell  \, \mbox{\rm Re} [(A_1 + C_1) (B_1^\ast + D_1^\ast)] \nnb \\
&-& \frac{1}{r} m_{B_s}^2 m_\ell (1+3 r - s)  \,
    \Big[ \mbox{\rm Re}(B_1 D_2^\ast) -  \mbox{\rm Re}(B_2 D_1^\ast)\Big] \nnb \\
&-&\frac{1}{r s} m_\ell (1- r - s)
    \Big[ \vel B_1 \ver^2 - \vel D_1 \ver^2\Big] \nnb \\
&+&\frac{1}{r s} (2 m_\ell^2  - m_{B_s}^2 s ) (1- r - s)
    \Big[ \mbox{\rm Re}(B_1 B_5^\ast) - \mbox{\rm Re}(D_1 B_4^\ast)\Big] \nnb \\
&+&\frac{1}{r} m_{B_s}^2 m_\ell (1- r - s) \,
    \mbox{\rm Re} [(B_1 + D_1) (B_3^\ast - D_3^\ast)] \nnb \\
&-&\frac{1}{r s} m_{B_s}^2 \lambda (2 m_\ell^2  - m_{B_s}^2 s )
    \Big[  \mbox{\rm Re}(B_2 B_5^\ast) -  \mbox{\rm Re}(D_2 B_4^\ast)\Big] \nnb \\
&-&\frac{1}{r s} m_{B_s}^4 m_\ell(1-r) \lambda
    \Big[ \vel B_2 \ver^2 - \vel D_2 \ver^2\Big]
    - \frac{1}{r} m_{B_s}^4 m_\ell \lambda  \,
    \mbox{\rm Re} [(B_2 + D_2) (B_3^\ast - D_3^\ast)] \nnb \\
&+&\frac{1}{r s} m_{B_s}^2 m_\ell [\lambda + (1-r-s) ( 1-r)]
    \Big[\mbox{\rm Re}(B_1 B_2^\ast) - \mbox{\rm Re}(D_1 D_2^\ast)\Big] \nnb \\
&+&\frac{2}{r s}  m_\ell^2 (1-r-s)
    \Big[\mbox{\rm Re}(B_1 B_4^\ast)-\mbox{\rm Re}(D_1 B_5^\ast)\Big] \nnb \\
&+&\frac{2}{r s} m_{B_s}^2  m_\ell^2 \lambda
    \Big[\mbox{\rm Re}(D_2 B_5^\ast) - \mbox{\rm Re}(B_2 B_4^\ast)\Big] \nnb \\
&+&\frac{16}{r s} \lambda m_{B_s}^2 m_\ell^2
    \,\mbox{\rm Re}[(B_1-D_1) (B_7 C_{TE})^\ast] \nnb \\
&-&\frac{16}{r s} \lambda m_{B_s}^4 m_\ell^2 (1-r)
    \,\mbox{\rm Re}[(B_2-D_2) (B_7 C_{TE})^\ast] \nnb \\
&-&\frac{8}{r} \lambda m_{B_s}^4 m_\ell
    \,\mbox{\rm Re}[(B_4-B_5) (B_7 C_{TE})^\ast] \nnb \\
&-&\frac{16}{r} \lambda m_{B_s}^4 m_\ell^2
    \,\mbox{\rm Re}[(B_3-D_3) (B_7 C_{TE})^\ast] \nnb \\
&-&\frac{32}{r s} m_\ell^2 (1-r-s)
    \,\mbox{\rm Re}[(B_1-D_1) (B_6 C_{TE})^\ast] \nnb \\
&+&\frac{32}{r s} m_{B_s}^2 m_\ell^2 (1-r) (1-r-s)
    \,\mbox{\rm Re}[(B_2-D_2) (B_6 C_{TE})^\ast] \nnb \\
&+&\frac{16}{r} m_{B_s}^2 m_\ell  (1-r-s)
    \,\mbox{\rm Re}[(B_4-B_5) (B_6 C_{TE})^\ast] \nnb \\
&+&\frac{32}{r} m_{B_s}^2 m_\ell^2  (1-r-s)
    \,\mbox{\rm Re}[(B_3-D_3) (B_6 C_{TE})^\ast] \nnb \\
&+& 16 m_{B_s}^2  \Big(
    4 m_\ell^2 \, \mbox{\rm Re}[A_1^\ast (C_T-2 C_{TE}) B_6]
    - m_{B_s}^2 s \, \mbox{\rm Re}[A_1^\ast (C_T+2 C_{TE}) B_6] \Big)\nnb \\
&-& 16 m_{B_s}^2  \Big(
    4 m_\ell^2 \, \mbox{\rm Re}[C_1^\ast (C_T+2 C_{TE}) B_6]
    - m_{B_s}^2 s \, \mbox{\rm Re}[C_1^\ast (C_T-2 C_{TE}) B_6] \Big)\nnb \\
&-& \frac{32}{s} m_{B_s}^2 (1-r) \Big(
    4 m_\ell^2 \, \mbox{\rm Re}[A_1^\ast (C_T-2 C_{TE}) T_1]
    - m_{B_s}^2 s \, \mbox{\rm Re}[A_1^\ast (C_T+2 C_{TE}) T_1] \Big)\nnb \\
&+& \frac{32}{s} m_{B_s}^2 (1-r) \Big(
    4 m_\ell^2 \, \mbox{\rm Re}[C_1^\ast (C_T+2 C_{TE}) T_1]
    - m_{B_s}^2 s \, \mbox{\rm Re}[C_1^\ast (C_T-2 C_{TE}) T_1] \Big)\nnb \\
&-&\frac{64}{r s} m_{B_s}^2 m_\ell^2 (1-r) (1+3 r-s)
    \,\mbox{\rm Re}[(B_2- D_2) (T_1 C_{TE})^\ast] \nnb \\
&-&\frac{64}{r} m_{B_s}^2 m_\ell^2  (1+3 r-s)
    \,\mbox{\rm Re}[(B_3-D_3) (T_1 C_{TE})^\ast] \nnb \\
&-&\frac{32}{r} m_{B_s}^2 m_\ell  (1+3 r-s)
    \,\mbox{\rm Re}[(B_4-B_5) (T_1 C_{TE})^\ast] \nnb \\
&-&\frac{64}{r s} [m_{B_s}^2 r s - m_\ell^2 (1+7 r -s)]
    \,\mbox{\rm Re}[(B_1-D_1) (T_1 C_{TE})^\ast] \nnb \\
&-& \frac{32}{s} (4 m_\ell^2 + m_{B_s}^2 s)
    \,\mbox{\rm Re}[(B_1+D_1) (T_1 C_T)^\ast] \nnb \\
&-&2048  m_{B_s}^2 m_\ell
    \,\mbox{\rm Re}[(C_T g)(B_6 C_{TE})^\ast] \nnb \\
&+&\frac{4096}{s} m_{B_s}^2 m_\ell (1-r) \vel T_1 \ver^2
    \, \mbox{\rm Re}(C_T C_{TE}^\ast)
    \Bigg\}~.
\eea

The normal polarization $P_N^-$ for $\ell^-$
\bea \label{pnm}
    P_N^-&=& \frac{1}{\Delta} \pi v m_{B_s}^3 \sqrt{s \lambda} \Bigg\{
    8 m_\ell \, \mbox{\rm Im}[(B_1^\ast C_1) + (A_1^\ast D_1)] \nnb \\
&-& \frac{1}{r} m_{B_s}^2 \lambda
    \,\mbox{\rm Im}[(B_2^\ast B_4) + (D_2^\ast B_5)] \nnb \\
&+& \frac{1}{r} m_{B_s}^2 m_\ell \lambda
    \,\mbox{\rm Im}[(B_2-D_2) (B_3^\ast-D_3^\ast)] \nnb \\
&-&\frac{1}{r} m_\ell \,(1 + 3 r - s)
    \,\mbox{\rm Im}[(B_1-D_1) (B_2^\ast-D_2^\ast)] \nnb \\
&+&\frac{1}{r} (1 - r - s)
    \, \mbox{\rm Im}[(B_1^\ast B_4) + (D_1^\ast B_5)] \nnb \\
&-&\frac{1}{r} m_\ell \,(1 - r - s)
    \,\mbox{\rm Im}[(B_1-D_1) (B_3^\ast-D_3^\ast)]  \nnb \\
&-&\frac{8}{r} m_{B_s}^2 m_\ell \lambda
    \,\mbox{\rm Im}[(B_4+B_5)(B_7 C_{TE})^\ast] \nnb \\
&+& \frac{16}{r} m_\ell \,(1-r-s)
    \,\mbox{\rm Im}[(B_4+B_5)(B_6 C_{TE})^\ast] \nnb \\
&-& \frac{32}{r} m_\ell \,(1+3 r-s)
    \,\mbox{\rm Im}[(B_4+B_5)(T_1 C_{TE})^\ast] \nnb \\
&-& 16 m_{B_s}^2 s \Big(
    \,\mbox{\rm Im}[A_1^\ast (C_T-2 C_{TE}) B_6] +
    \mbox{\rm Im}[C_1^\ast (C_T+2 C_{TE}) B_6] \Big) \nnb \\
&+& 32 m_{B_s}^2 (1-r) \Big(
    \,\mbox{\rm Im}[A_1^\ast (C_T-2 C_{TE}) T_1] +
    \mbox{\rm Im}[C_1^\ast (C_T+2 C_{TE}) T_1] \Big) \nnb \\
&+& 32 \Big(
    \mbox{\rm Im}[B_1^\ast (C_T-2 C_{TE}) T_1]
    - \mbox{\rm Im}[D_1^\ast (C_T+2 C_{TE}) T_1] \Big) \nnb \\
&+& 512 m_\ell \,
    \ga \vel C_T \ver^2 - 4 \vel C_{TE} \ver^2 \dr
    \,\mbox{\rm Im}(B_6^\ast T_1)
    \Bigg\}~,
\eea
and   $P_N^+$ for $\ell^+$
\bea \label{pnp}
    P_N^+&=& \frac{1}{\Delta} \pi v m_{B_s}^3 \sqrt{s \lambda} \Bigg\{
    - 8 m_\ell \, \mbox{\rm Im}[(B_1^\ast C_1) + (A_1^\ast D_1)] \nnb \\
&+& \frac{1}{r} m_{B_s}^2 \lambda
    \,\mbox{\rm Im}[(B_2^\ast B_4) + (D_2^\ast B_5)] \nnb \\
&+& \frac{1}{r} m_{B_s}^2 m_\ell \lambda
    \,\mbox{\rm Im}[(B_2-D_2) (B_3^\ast-D_3^\ast)] \nnb \\
&-&\frac{1}{r} m_\ell \,(1 + 3 r - s)
    \,\mbox{\rm Im}[(B_1-D_1) (B_2^\ast-D_2^\ast)] \nnb \\
&-&\frac{1}{r} (1 - r - s)
    \, \mbox{\rm Im}[(B_1^\ast B_5) + (D_1^\ast B_4)] \nnb \\
&-&\frac{1}{r} m_\ell \,(1 - r - s)
    \,\mbox{\rm Im}[(B_1-D_1) (B_3^\ast-D_3^\ast)]  \nnb \\
&+&\frac{8}{r} m_{B_s}^2 m_\ell \lambda
    \,\mbox{\rm Im}[(B_4+B_5)(B_7 C_{TE})^\ast] \nnb \\
&-& \frac{16}{r} m_\ell \,(1-r-s)
    \,\mbox{\rm Im}[(B_4+B_5)(B_6 C_{TE})^\ast] \nnb \\
&+& \frac{32}{r} m_\ell \,(1+3 r-s)
    \,\mbox{\rm Im}[(B_4+B_5)(T_1 C_{TE})^\ast] \nnb \\
&-& 16 m_{B_s}^2 s \Big(
    \,\mbox{\rm Im}[A_1^\ast (C_T+2 C_{TE}) B_6] +
    \mbox{\rm Im}[C_1^\ast (C_T-2 C_{TE}) B_6] \Big) \nnb \\
&+& 32 m_{B_s}^2 (1-r) \Big(
    \,\mbox{\rm Im}[A_1^\ast (C_T+2 C_{TE}) T_1] +
    \mbox{\rm Im}[C_1^\ast (C_T-2 C_{TE}) T_1] \Big) \nnb \\
&-& 32 \Big(
    \mbox{\rm Im}[B_1^\ast (C_T+2 C_{TE}) T_1]
    - \mbox{\rm Im}[D_1^\ast (C_T-2 C_{TE}) T_1] \Big) \nnb \\
&+& 512 m_\ell \,
    \ga \vel C_T \ver^2 - 4 \vel C_{TE} \ver^2 \dr
    \,\mbox{\rm Im}(B_6^\ast T_1)
    \Bigg\}~.
\eea
The combined longitudinal polarization $P_L^- + P_L^+$, from (\ref{plm}):
\bea \label{lpl} P_L^- + P_L^+ &=& \frac{4}{\Delta}  \,m_{B_s}^2 v
\,\Bigg\{
    \frac{2}{r} m_\ell \lambda\,
    \mbox{\rm Re} [(B_1 - D_1) (B_4^\ast + B_5^\ast)] \nnb \\
&-& \frac{2}{r} m_{B_s}^2 m_\ell \lambda (1-r) \,
    \mbox{\rm Re} [(B_2 - D_2) (B_4^\ast + B_5^\ast)] \nnb \\
&-&\frac{1}{r} m_{B_s}^2 s \lambda
    \Big( \vel B_4 \ver^2 - \vel B_5 \ver^2\Big) -
    \frac{2}{r} m_{B_s}^2 m_\ell s \lambda \,
    \mbox{\rm Re} [(B_3 - D_3) (B_4^\ast + B_5^\ast)] \nnb \\
&+& \frac{8}{3 r} m_{B_s}^4 m_\ell \lambda^2 \,
    \mbox{\rm Re} [(B_2 + D_2) (B_7 C_T)^\ast] \nnb \\
&+& \frac{32}{3 r} m_{B_s}^6 s  \lambda^2 \vel B_7 \ver^2
    \mbox{\rm Re} (C_T C_{TE}^\ast) \nnb \\
&-& \frac{8}{3 r} m_{B_s}^2 m_\ell \lambda (1-r-s) \,
    \mbox{\rm Re} [(B_1 + D_1) (B_7 C_T)^\ast] \nnb \\
&-& \frac{16}{3 r} m_{B_s}^2 m_\ell \lambda (1-r-s) \,
    \mbox{\rm Re} [(B_2 + D_2) (B_6 C_T)^\ast] \nnb \\
&-& \frac{128}{3 r} m_{B_s}^4 s \lambda(1-r-s) \,
    \mbox{\rm Re} (B_6 B_7^\ast) \, \mbox{\rm Re} (C_T C_{TE}^\ast) \nnb \\
&+& \frac{16}{3 r}  m_\ell (\lambda + 12 r s) \,
    \mbox{\rm Re} [(B_1 + D_1) (B_6 C_T)^\ast] \nnb \\
&+& \frac{128}{3 r} m_{B_s}^2 s (\lambda + 12 r s) \,
    \vel B_6 \ver^2 \mbox{\rm Re} (C_T C_{TE}^\ast) \nnb \\
&+& \frac{512}{3 r} m_{B_s}^2 \, [ \lambda (4 r + s) + 12 r (1-r)^2
]
    \, \vel T_1 \ver^2 \,\mbox{\rm Re} (C_T C_{TE}^\ast)\nnb \\
&-& \frac{512}{3 r} m_{B_s}^2 s \, [\lambda +12 r (1-r) ]\,
    \mbox{\rm Re} (T_1 B_6^\ast) \, \mbox{\rm Re} (C_T C_{TE}^\ast) \nnb \\
&+& \frac{256}{3 r} m_{B_s}^4 s \lambda (1+3r -s) \,
    \mbox{\rm Re} (T_1 B_7^\ast) \, \mbox{\rm Re} (C_T C_{TE}^\ast) \nnb \\
&+& \frac{512}{3} m_{B_s}^2 m_\ell \lambda \,
    \mbox{\rm Re} [(A_1 + C_1) (T_1 C_{TE})^\ast] \nnb \\
&-& \frac{32}{3 r}  m_\ell  \, [\lambda +12 r (1-r) ]\,
    \mbox{\rm Re} [(B_1 + D_1) (T_1 C_T)^\ast] \nnb \\
&+& \frac{32}{3 r} m_{B_s}^2 m_\ell  \lambda (1+3r -s) \,
    \mbox{\rm Re} [(B_2 + D_2) (T_1 C_T)^\ast]
    \Bigg\}~.
\eea
The combined transversal polarization $P_T^- - P_T^+$, from (\ref{ptm}) and
(\ref{ptp}):
\bea \label{tmt} P_T^- - P_T^+ &=& \frac{\pi}{\Delta} m_{B_s}
    \sqrt{s \lambda} \Bigg\{ \frac{2}{r s} m_{B_s}^4 m_\ell (1-r)
    \lambda
    \Big[ \vel B_2 \ver^2 - \vel D_2 \ver^2\Big]\nnb \\
&+& \frac{1}{r} m_{B_s}^4 \lambda \,
    \mbox{\rm Re} [(B_2 + D_2) (B_4^\ast - B_5^\ast)] \nnb \\
&+& \frac{2}{r} m_{B_s}^4 m_\ell \lambda \,
    \mbox{\rm Re} [(B_2 + D_2) (B_3^\ast - D_3^\ast)] \nnb \\
&+& \frac{2}{r} m_{B_s}^2 m_\ell (1+3 r - s)  \,
    \Big[ \mbox{\rm Re}(B_1 D_2^\ast) -  \mbox{\rm Re}(B_2 D_1^\ast)\Big] \nnb \\
&+& \frac{2}{rs} m_\ell (1-r-s)
    \Big[ \vel B_1 \ver^2 - \vel D_1 \ver^2\Big]\nnb \\
&-& \frac{1}{r} m_{B_s}^2 (1-r-s)
    \mbox{\rm Re} [(B_1 + D_1) (B_4^\ast - B_5^\ast)] \nnb \\
&-& \frac{2}{r} m_{B_s}^2 m_\ell (1-r-s)
    \mbox{\rm Re} [(B_1 + D_1) (B_3^\ast - D_3^\ast)] \nnb \\
&-&\frac{2}{r s} m_{B_s}^2 m_\ell [\lambda + (1-r) (1-r-s)]
    \Big[\mbox{\rm Re}(B_1 B_2^\ast) - \mbox{\rm Re}(D_1 D_2^\ast)\Big] \nnb \\
&-&\frac{32}{r s} m_{B_s}^2 m_\ell^2 \lambda \,
    \mbox{\rm Re}[(B_1-D_1) (B_7 C_{TE})^\ast] \nnb \\
&+&\frac{32}{r s} m_{B_s}^4 m_\ell^2 \lambda (1-r)\,
    \mbox{\rm Re}[(B_2-D_2) (B_7 C_{TE})^\ast] \nnb \\
&+&\frac{16}{r} m_{B_s}^4 m_\ell \lambda \,
    \mbox{\rm Re}[(B_4-B_5) (B_7 C_{TE})^\ast] \nnb \\
&+&\frac{32}{r} m_{B_s}^4 m_\ell^2 \lambda \,
    \mbox{\rm Re}[(B_3-D_3) (B_7 C_{TE})^\ast] \nnb \\
&+&\frac{64}{r s} m_\ell^2 (1-r-s) \,
    \mbox{\rm Re}[(B_1-D_1) (B_6 C_{TE})^\ast] \nnb \\
&-&\frac{64}{r s} m_{B_s}^2 m_\ell^2 (1-r)(1-r-s) \,
    \mbox{\rm Re}[(B_2-D_2) (B_6 C_{TE})^\ast] \nnb \\
&-&\frac{32}{r} m_{B_s}^2 m_\ell (1-r-s) \,
    \mbox{\rm Re}[(B_4-B_5) (B_6 C_{TE})^\ast] \nnb \\
&-&\frac{64}{r} m_{B_s}^2 m_\ell^2 (1-r-s) \,
    \mbox{\rm Re}[(B_3-D_3) (B_6 C_{TE})^\ast] \nnb \\
&+& 32 m_{B_s}^4 s v^2 \,
    \mbox{\rm Re}[(A_1-C_1) (B_6 C_T)^\ast] \nnb \\
&+&\frac{64}{r} m_{B_s}^2 m_\ell (1+3 r-s) \,
    \mbox{\rm Re}[(B_4-B_5) (T_1 C_{TE})^\ast] \nnb \\
&-& 64 m_{B_s}^4 (1-r) v^2 \,
    \mbox{\rm Re}[(A_1-C_1) (T_1 C_T)^\ast] \nnb \\
&+&\frac{128}{r s} [m_{B_s}^2 r s - m_\ell^2 (1+7 r-s)] \,
    \mbox{\rm Re}[(B_1-D_1) (T_1 C_{TE})^\ast] \nnb \\
&+&\frac{128}{r s} m_{B_s}^2 m_\ell^2 (1-r) (1+3 r-s)
    \mbox{\rm Re}[(B_2-D_2) (T_1 C_{TE})^\ast] \nnb \\
&+&\frac{128}{r} m_{B_s}^2 m_\ell^2 (1+3 r-s)
    \mbox{\rm Re}[(B_3-D_3) (T_1 C_{TE})^\ast]
    \Bigg\}
\eea

The combined normal polarization $P_N^- + P_N^+$,  from (\ref{pnm}) and
(\ref{pnp}):
\bea \label{npn}
    P_N^- + P_N^+ &=& \frac{1}{\Delta} \pi v m_{B_s}^3 \sqrt{s\lambda} \Bigg\{
    - \frac{2}{r} m_\ell (1+3 r -s) \,
    \mbox{\rm Im} [(B_1 - D_1) (B_2^\ast - D_2^\ast)] \nnb \\
&-& \frac{2}{r} m_\ell (1-r -s) \,
    \mbox{\rm Im} [(B_1 - D_1) (B_3^\ast - D_3^\ast)] \nnb \\
&-& \frac{1}{r} (1-r -s) \,
    \mbox{\rm Im} [(B_1 - D_1) (B_4^\ast - B_5^\ast)] \nnb \\
&+&\frac{2}{r} m_{B_s}^2  m_\ell \lambda \,
    \mbox{\rm Im} [(B_2 - D_2) (B_3^\ast - D_3^\ast)] \nnb \\
&+&\frac{1}{r} m_{B_s}^2 \lambda \,
    \mbox{\rm Im} [(B_2 - D_2) (B_4^\ast - B_5^\ast)] \nnb \\
&+& 32 m_{B_s}^2 s \, \mbox{\rm Im} [(A_1 + C_1)(B_6 C_T)^\ast] \nnb \\
&+&1024 m_\ell \Big(  \vel C_T \ver^2  - \vel 4 C_{TE} \ver^2  \Big)
    \mbox{\rm Im} (B_6^\ast T_1) \nnb \\
&-& 64 m_{B_s}^2 (1-r) \, \mbox{\rm Im} [(A_1 + C_1)(T_1 C_T)^\ast] \nnb \\
&+& 128 \, \mbox{\rm Im} [(B_1 + D_1)(T_1 C_{TE})^\ast]
    \Bigg\}
\eea

\newpage
\renewcommand{\topfraction}{.99}
\renewcommand{\bottomfraction}{.99}
\renewcommand{\textfraction}{.01}
\renewcommand{\floatpagefraction}{.99}

\clearpage
\clearpage
\clearpage
\clearpage
\clearpage
\clearpage
\end{document}